\documentclass[a4paper, 11pt]{article}
\usepackage{lipsum}
\pdfoutput=1
\usepackage{comment}
\usepackage{jcappub} 
\usepackage{graphicx}
\usepackage{float}
\usepackage{xcolor}
\usepackage[colorinlistoftodos]{todonotes}

\usepackage[normalem]{ulem}

\usepackage[T1]{fontenc}
\usepackage{subfig}
\usepackage{booktabs}
\usepackage{hyperref}

\title{\boldmath Primordial feature constraints from BOSS $+$ eBOSS}

\author[a]{Thiago Mergulh\~ao,}
\author[a]{Florian Beutler,}
\author[a]{John A. Peacock.}

\affiliation[a]{\small Institute for Astronomy, University of Edinburgh, Royal Observatory, Blackford Hill, Edinburgh EH9 3HJ, UK}

\emailAdd{thiago.mergulhao@ed.ac.uk}

\abstract{Understanding the universe in its pristine epoch is crucial in order to obtain a concise comprehension of the late-time universe. Although current data in cosmology are compatible with Gaussian primordial perturbations whose power spectrum follows a nearly scale-invariant power law, this need not be the case when a fundamental theoretical construction is assumed. These extended models lead to sharp features in the primordial power spectrum, breaking its scale invariance. In this work, we obtain combined constraints on four primordial feature models by using the final data release of the BOSS galaxies and eBOSS quasars. By pushing towards the fundamental mode of these surveys and using the larger eBOSS volume, we were able to extend the feature parameter space (i.e. the feature frequency $\omega$) by a factor of four compared to previous analyses using BOSS. While we did not detect any significant features, previous work showed that next-generation galaxy surveys such as DESI will improve the sensitivity to features by a factor of 7, and will also extend the parameter space by a factor of 2.5.}

\begin{document}

\maketitle
\flushbottom

\section{Introduction}
\label{sec: introduction}
Over the last few decades, we have witnessed major steps in cosmological physics toward a better understanding of the universe and its nature. Projects such as the Planck collaboration \cite{aghanim2020planck, akrami2020planck}, which probed the cosmic microwave background (CMB) in an unprecedented way, and the Baryon Oscillation Spectroscopic Survey (BOSS), which constructed a three-dimensional map of the late universe distribution of galaxies, imposed strong constraints on the standard cosmological model. However, there are still questions that need to be addressed, many of them regarding the early universe and its dynamics -- the same desideratum present in particle physics to look for a more fundamental description of nature is also present in cosmology, and in some sense, they are fairly connected \cite{Achucarro:2022qrl}.

The primordial power spectrum of curvature perturbations is the seed for structure formation, and its form depends on the physical processes that operate at early times. In the standard model of cosmology, these perturbations are: adiabatic; their statistical distribution is highly Gaussian; and their power spectrum is a nearly scale-invariant power-law. Two of the six free parameters in the standard model are associated with this: the amplitude of the scalar perturbations, $A_{\rm s}$, and the spectral index $n_{\rm s}$. Although this simple model allows us to explain a vast number of observables, there are many reasons why this model by itself is not natural, and there are even more ways to construct other viable scenarios. Among these, we give the following examples\footnote{See \cite{chluba2015features} for a recent review about these and other primordial feature models.}:
\begin{itemize}
    \item Embedding inflation in grand unified theories can result in a series of phase transitions \cite{Vilenkin:1982wt} that can induce steps in the potential \cite{Adams:1997de, Adams:2001vc}, a mechanism that gives a sudden burst to the inflaton's kinetic energy. The impact of these steps has been extensively analysed in the literature and they usually cause localized oscillatory signals in the power spectrum \cite{miranda2014inflationary, Dvorkin:2009ne, Hu:2011vr, Adshead:2011jq}.
    \item The inflaton can be only one degree of freedom in a multi-field scenario. Considering this framework, the inflaton can go through a region in field space where some of the massive degrees of freedom to which it is coupled suddenly become massless, causing the production of heavy field quanta \cite{Barnaby:2009mc, Chung:1999ve, Dvali:2003ar, Kofman:2004yc, Romano:2008rr, Green:2009ds}. Another scenario is when any of these other hidden fields perform a sudden turn in their configuration space, which produces oscillating features on top of the usual power-law prediction \cite{KOFMAN1988508, Braglia:2020fms, Price:2014xpa, Gao:2012uq, Achucarro:2010da}.
    \item Inflation is known to magnify quantum fluctuations generated in the indefinite past by many orders of magnitude, making it relevant for later cosmic evolution. The initial wavelength of these fluctuations can be as small as the Planck scale, and transPlanckian physics can modify the primordial power spectrum, including global features \cite{Danielsson:2002kx,Martin:2003kp, Bozza:2003pr, Chen:2011zf, Fergusson:2014tza, Fergusson:2014hya}.
    \item Logarithmically oscillating features can be generated in both natural \cite{Freese:1990rb} and axion monodromy inflation \cite{McAllister:2008hb}. Given the importance of axions in cosmology \cite{Kawasaki:2013ae} and the simplicity of their associated corrections to the primordial power spectrum, constraining their impact with current and forthcoming data may be vital for studying their existence.
\end{itemize}
The physical mechanism behind these primordial features (PF) is quite broad and intrinsically connected to our fundamental descriptions of nature. Hence, any detection of an additional inflationary signal lurking in data could drastically change our picture of the early phases of the universe. Consequently, many groups have been trying to understand the impact of these primordial features on data and to develop pipelines to detect them. There is already a plethora of analyses using WMAP \cite{2010JCAP...04..010H, Shafieloo:2006hs, 2012MNRAS.421..369M, Martin:2003sg, Martin:2004yi, Flauger:2009ab, Aich:2011qv, Peiris:2013opa} and Planck \cite{miranda2014inflationary, Easther:2013kla, akrami2020planck, Miranda:2013wxa, Braglia:2021rej, Meerburg:2013dla, Chen:2014joa} datasets. Forecasts for future CMB surveys are present in recent works \cite{2022JCAP...09..024C, Hamann:2022apw, Braglia:2022ftm}. Gravitational waves are one of the main byproducts of inflation; the presence of PF could also affect the stochastic gravitational wave background \cite{2022JHEP...03..196F, Fumagalli:2020nvq, Braglia:2020taf, Fumagalli:2021cel, Witkowski:2021raz, Fumagalli:2021mpc}, and \cite{Fumagalli:2021dtd} showed a forecast for LISA. It is important to mention that even if inflation is not assumed to be true, PF is one of the main signals that could potentially help us to distinguish inflation from other scenarios \cite{Chen:2011zf, 2011PhLB..706..111C, Chen:2012ja, Chen:2014joa}. 

Additional primordial physics can affect the clustering of tracers, which makes large-scale structure (LSS) a potential source of information on PF to complement other datasets. These signals, if present, are susceptible to non-linear effects arising from late-time structure formation. A well-known example of this effect is the degradation of the Baryon Acoustic Oscillation (BAO) peak due to non-linear clustering \cite{Crocce:2007dt}. Nevertheless, much effort has been made to address this problem. In \cite{Vlah:2015zda, Vasudevan:2019ewf, beutler2019primordial, Chen:2020ckc, Ballardini:2019tuc} the impact of non-linearities on PF was studied, providing analytical predictions using perturbation theory. These results were tested with N-body simulations, proving to be precise enough for analysis using current and future datasets. In addition, by using perturbation theory one can partially undo the non-linear effects by performing a field reconstruction \cite{Shi:2017gqs, Ota:2021caz, Seo:2021nev}. Altogether, this allowed the development of a robust analysis of PF with LSS \cite{beutler2019primordial}; this showed that the final BOSS data release improved on the Planck constraints on both linear and logarithmic features over a considerable volume in parameter space.

 The importance of LSS constraints on primordial features and the fast progression of stage-IV spectroscopic surveys like DESI and Euclid motivate us to start developing  an in-depth search for PF using large-scale structure data. In this work, we push further the analysis performed in \cite{beutler2019primordial} and obtain combined PF credible intervals from the final BOSS galaxies and eBOSS quasars. We show that the high number density of the former along with the greater volume of the latter are complementary, allowing us to obtain credible intervals for primordial features over a region in parameter space never explored before. This work is structured in the following way: In Section \ref{sec: clustering_template} we explain how the feature signals can be probed by a broadband analysis of the galaxy power spectrum. In Section \ref{sec: primordial_features} we show all the primordial feature signals that will be considered in this work and discuss their impact on the distribution of galaxies. In Section \ref{sec: data_used} we explain the data we used. In Section \ref{sec: methods} we scrutinize our methodology along with the statistical tools used to check the significance of our results, which is presented in Section \ref{sec:results}.

\section{Galaxy clustering template}
The presence of primordial features changes the initial power spectrum of curvature perturbations. In this section, we show how it is possible to propagate the primordial feature signal into the evolved galaxy field at linear order.
\label{sec: clustering_template}

\subsection{Propagating the primordial feature to the evolved matter field}
The primordial feature signals cannot be mimicked by any specific combination of the cosmological parameters in the standard model. For this reason, we follow \cite{beutler2019primordial} and assume that as a first approximation, one can simply marginalize over the broadband signal in the galaxy power spectrum and then look for residual signals. To see how PF propagate to the evolved matter field we consider the linear evolution of the initial conditions through the transfer function, $T(k)$:
\begin{equation}
    \label{eqn: linear_evolution}
    P(z,k)=k^{4} T(k)^{2} D(z)^{2} P_{\mathcal{R}}(k)\;,
\end{equation}
where $D(z)$ is the linear growth factor and $P_{\mathcal{R}}$ is the primordial power spectrum. In the simplest models of inflation, this power spectrum is given by
\begin{equation}
    \label{eqn: P_R(0)}
    P_{\mathcal{R}, 0}(k)=\frac{2 \pi^{2}}{k^{3}} \mathcal{P}_{\mathcal{R}, 0}(k)=\frac{2 \pi^{2} A_{\mathrm{s}}}{k^{3}}\left(\frac{k}{k_{\star}}\right)^{n_{\mathrm{s}}-1}\;.
\end{equation}
The contributions of primordial features, $\delta P_{\mathcal{R}}$, show up as corrections to Eq.\,(\ref{eqn: P_R(0)}) in the form
\begin{equation}
    \label{eqn: P_R}
    P_{\mathcal{R}}(k)=P_{\mathcal{R}, 0}(k)\left[1+\delta P_{\mathcal{R}}(k)\right]\;.
\end{equation}
By plugging Eq.\,(\ref{eqn: P_R}) into Eq.\,(\ref{eqn: linear_evolution}) it is clear that
\begin{eqnarray}
\nonumber
P(z,k) &=& P_{\rm lin}(z,k)\left[1+\delta P_{\mathcal{R}}(k)\right]\\ \label{eqn: matter_with_feature}
&=& P^{\rm nw}_{\rm lin}(z,k) + P^{\rm w}_{\rm lin BAO}(z,k) + \delta P_{\mathcal{R}}P^{\rm nw}_{\rm lin} + P^{\rm w}_{\rm lin, BAO}\delta P_{\mathcal{R}}(k)\;,
\end{eqnarray}
where we split the usual linear matter field into a smooth part and the BAO contributions, $P_{\rm lin} = P^{\rm nw}_{\rm lin} + P^{\rm w}_{\rm lin, BAO}$. Eq.\,(\ref{eqn: matter_with_feature}) is our final result for the evolved matter power spectrum with PF.

\subsection{Galaxy clustering with primordial features}

The galaxy power spectrum is introduced in the following way. In equation Eq.\,(\ref{eqn: matter_with_feature}) we factor out a linear smooth power spectrum, $P^{\rm nw}_{\rm lin}$, that contains both BAO and PF oscillations, $\delta P_{\mathcal{R}}$:
\begin{equation}
\label{eqn: matter_splitted}
    \tilde{P}(z,k) = P^{\rm nw}_{\rm lin}\left(1 + \mathcal{O}_{\rm lin} + \delta P_{\mathcal{R}} + \mathcal{O}_{\rm lin}\delta P_{\mathcal{R}}\right)\;,
\end{equation}
where $\mathcal{O}_{\rm lin}(k) \equiv P^{\rm w}_{\rm lin, BAO}/P^{\rm nw}_{\rm lin}$. We included a tilde in the expression above to emphasize that this is not the final result for the evolved power spectrum -- non-linear displacements will damp the oscillations and need to be taken into account (see \S\ref{sec: non_linear_damping}). Eq.\,(\ref{eqn: matter_splitted}) is the result for the matter field, and we need to connect it with our observable, the tracer field. Since we are interested only in modulations around the smooth power spectrum, we employ a phenomenological model instead of the usual bias expansion in perturbation theory: we assume that the connection between the matter and the galaxy field is given by broadband information encapsulated in a linear bias, $B$, a spectral distortion due to the non-linear velocity field, $F\left(k, \Sigma_s\right)$, and a series of broadband parameters, $A(k)$:  
\begin{equation}
\label{eqn: P_tilde}
    P^{\rm nw}_{\rm g}(k)=B^{2} P^{\mathrm{nw}}(k) F\left(k, \Sigma_{s}\right)+A(k)\;,
\end{equation}
with 
\begin{eqnarray}
\label{eqn: A(k)}
    A(k)&=&\frac{a_{1}}{k^{3}}+\frac{a_{2}}{k^{2}}+\frac{a_{3}}{k}+a_{4}+a_{5} k^{2}\;,\\
    \label{eqn: F(k)}
    F\left(k, \Sigma_{s}\right)&=&\frac{1}{\left(1+k^{2} \Sigma_{s}^{2} / 2\right)^{2}}\;.
\end{eqnarray}
The forms for $A(k)$ and $F(k, \Sigma_s)$ are well motivated and commonly used in clustering analysis \cite{Neveux:2020voa,Seo:2015eyw,BOSS:2016hvq,Ballinger:1996cd,Magira:1999bn, Park1994,Peacock:1993xg}.

\subsection{The non-linear damping of oscillatory features: infrared resummation}
\label{sec: non_linear_damping}
Although Eq.\,(\ref{eqn: matter_splitted}) considers the impact of both spectral distortion due to the velocity field and a set of broadband parameters to account for the broadband non-linear clustering that we marginalise over, there is still one effect that needs to be considered: the displacement of the galaxies due to long-wavelength perturbations. In Eulerian perturbation theory both the overdensity and velocity fields are considered to be small. However, since the usual time scale of the system is the Hubble time, $H(a)^{-1}$, the galaxies end up having significant displacements. In Eulerian perturbation theory, this displacement field is considered to be small, but this is not the case. Thus, the perturbations need to be resummed in order to have a consistent result \cite{Senatore:2014via}. In the case of a linear primordial feature, the effect of the damping will be the same as in the BAO case, and it transforms Eq.\,(\ref{eqn: matter_splitted}) into \cite{Vlah:2015zda, Vasudevan:2019ewf}:
\begin{equation}
\label{eqn: resummed_matter}
    P_{m}(k) = P^{\rm nw}(k)\left[1 + \left(\mathcal{O}_{\rm lin} + \delta P_{\mathcal{R}} + \mathcal{O}_{\rm lin}\delta P_{\mathcal{R}}\right)e^{-k^2\Sigma_{\rm lin}^2}\right]\;,
\end{equation}
where
\begin{equation}
    \Sigma_{\operatorname{lin}}^{2} \equiv \frac{1}{6 \pi^{2}} \int d q\,  P^{\mathrm{nw}}(q)\left[1-j_{0}\left(q \omega_{\operatorname{lin}}\right)+2 j_{2}\left(q \omega_{\operatorname{lin}}\right)\right]\;.
\end{equation}
Eq.\,(\ref{eqn: resummed_matter}) is the final result for the matter field after the large-scale bulk flow is considered. It can easily be extended to galaxy clustering:
\begin{eqnarray}
\label{eqn: common_template}
P_{g}(k)=P_{g}^{\mathrm{nw}}(k)\left\{1+[\mathcal{O}_{\rm lin}(k)+\delta P_{\mathcal{R}}^{X}(k)
+\mathcal{O}_{\rm lin}(k) \delta P_{\mathcal{R}}^{X}(k)] e^{-k^{2} \Sigma_{\mathrm{nl}}^{2} / 2}\right\}\;,
\end{eqnarray}
where the primordial feature corrections are represented by $\delta P_{\mathcal{R}}$. Before continuing, we emphasize that although the damping term included above, $e^{-k^{2} \Sigma_{\mathrm{nl}}^{2}/ 2}$, is well motivated for linear features, it is expected to have a scale-dependence for e.g. the logarithmic feature. In~\cite{beutler2019primordial}, the authors showed that as a first approximation, it is possible to use a scale-independent damping model for the logarithmic feature. Here we follow~\cite{beutler2019primordial} and approximate the non-linear damping by using the scale-independent parameter $\Sigma_{\rm nl}\approx \Sigma_{\rm lin}$. In \cite{Chen:2020ckc} the authors showed how the non-linear damping can be approximated analytically in perturbation theory for general sharp features, and we will employ their results in future work. The simple approximation that we make here has been tested in \cite{Ballardini:2019tuc} using data from an N-body simulation: they concluded that the template in Eq.~(\ref{eqn: common_template}) described their data reasonably well. This work considered dark matter only and it would be good to repeat it with a variety of galaxy-formation recipes; but our expression in Eq.~(\ref{eqn: common_template}) is essentially the one that is commonly used to describe damping of BAO features, so we do not expect its applicability to depend on such details.

Another point that needs to be addressed is the inclusion of the isotropic BAO scaling parameter, $\alpha$. In the standard analysis in Fourier space, this parameter specifies the frequency or wavelength of the BAO wiggles and shows up in Eq.\,(\ref{eqn: common_template}) as:
\begin{equation}
    \label{eqn: final_template}
P_{g}(k)=P_{g}^{\mathrm{nw}}(k)\{1+[\mathcal{O}_{\rm lin}(k/\alpha)+\delta P_{\zeta}^{X}(k)
+\mathcal{O}_{\rm lin}(k/\alpha) \delta P_{\zeta}^{X}(k)] e^{-k^{2} \Sigma_{\mathrm{nl}}^{2} / 2}\}\;.
\end{equation}
To summarize, in addition to the PF-free parameters, we also need to fit the following ten parameters per dataset:
\begin{equation}
    \{B_{\rm NGC},B_{\rm SGC}, a_1, a_2, a_3, a_4, a_5, \Sigma_s, \Sigma_{\rm lin}, \alpha\}\;.
    \label{eqn: common_params}
\end{equation}
Notice that we included two linear amplitudes, $B_{\rm NGC}$ and $B_{\rm SGC}$. The reason is that for a given dataset, we have two parts, the north galactic cap (NGC) and the south galactic cap (SGC). Therefore we allow them to have different linear amplitudes, but the other parameters are shared (see Section \ref{sec: data_used}). This model follows the standard BOSS BAO analysis \cite{BOSS:2016hvq}.

\section{Primordial feature signals}
\label{sec: primordial_features}


We now present more detail on the specific PF models that we will seek to constrain using LSS data.

\subsection{Linear features}

\begin{figure}[t!]
    \centering
        \includegraphics[width = 0.48\textwidth]{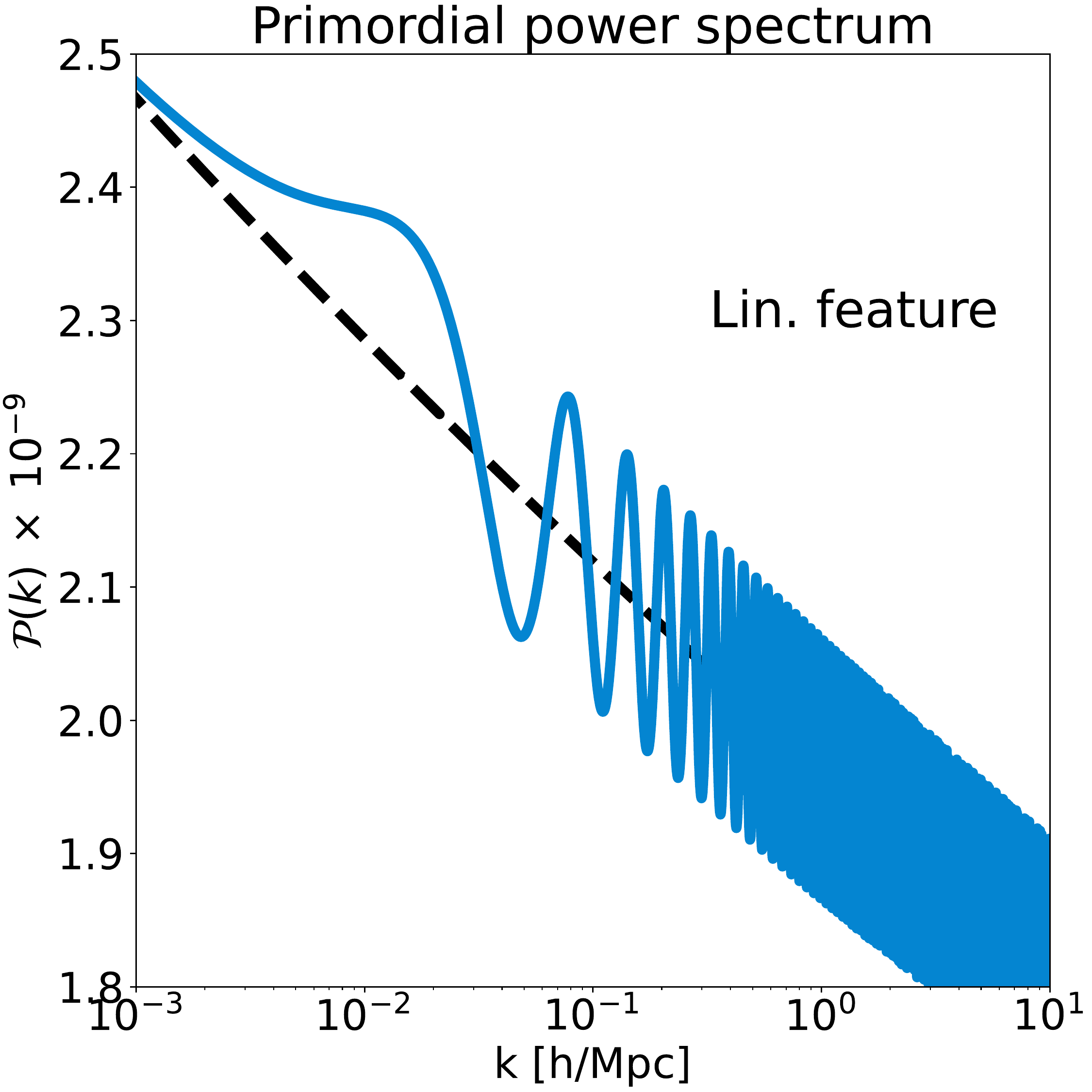}
    \includegraphics[width = 0.48\textwidth]{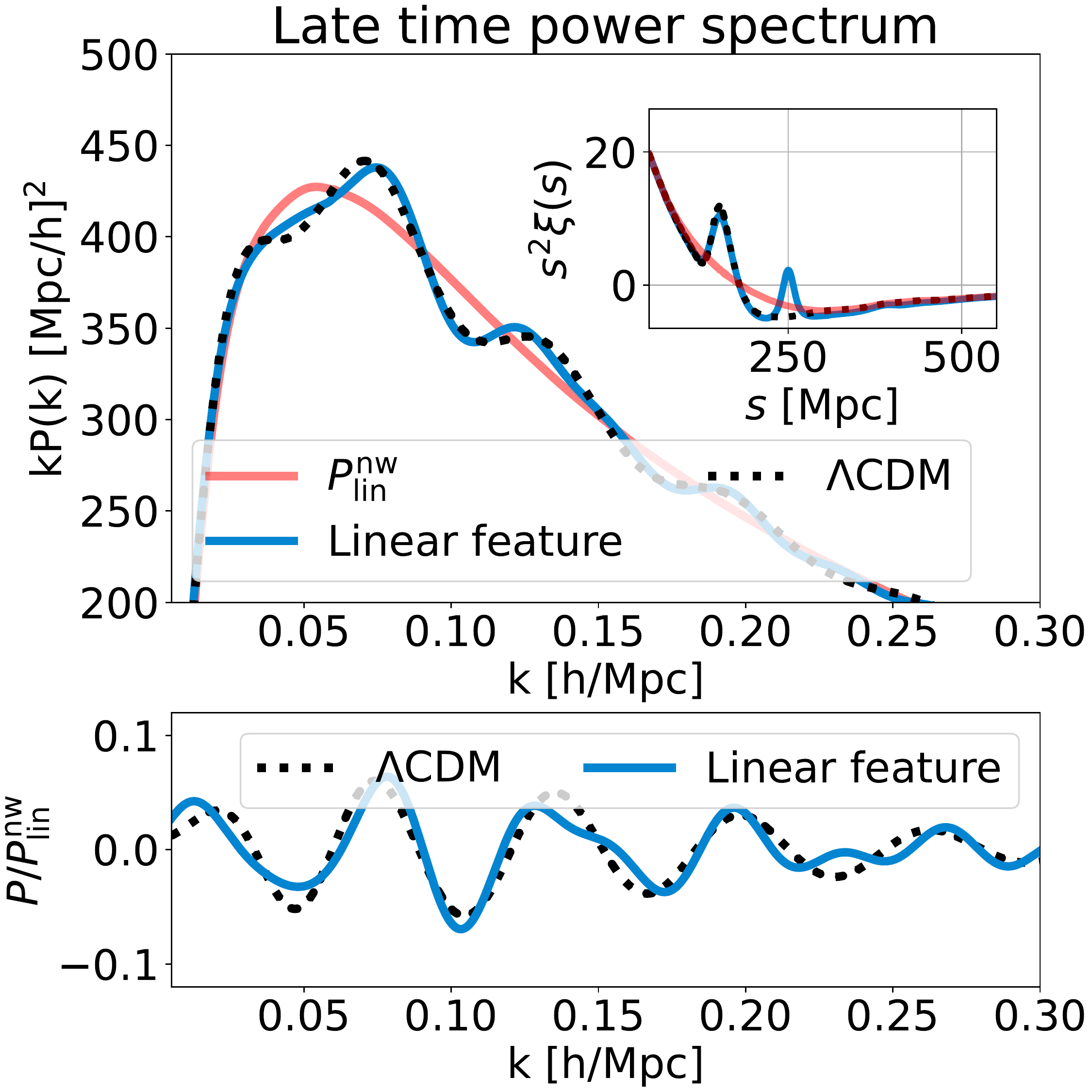}
    \caption{\textit{left:} The primordial curvature power spectrum in the vanilla model of inflation (black dashed line) and in the linear feature case (solid blue line). For this plot, we used $A_{\rm lin} = 0.05$, $\omega_{\rm lin} = 250\,\mathrm{Mpc}$, $\phi = 0$ and the BOSS high-$z$ window function; \textit{right:} An illustration of how the linear feature shows up in the evolved galaxy field.}
    \label{fig: linear_feature_plots}
\end{figure}

If some background inflationary quantities have a sudden change, the inflaton field can have a feature that is spread over its entire power spectrum. A general template for this signal is
\begin{equation}
\label{eqn: lin_feature}
    \delta \mathcal{P}_{\mathcal{R}} = A_{\rm lin} \sin(\omega_{\rm lin}k + \phi)\;,
\end{equation}
where $A_{\rm lin}$ is the feature amplitude, $\phi$ is some arbitrary phase and $\omega_{\rm lin}$ is the frequency. In configuration space, a linear feature appears as a second peak on the correlation function at a separation of $s = \omega_{\rm lin}$. This is shown in Figure \ref{fig: linear_feature_plots}. This kind of signal was already scanned for in both CMB \cite{akrami2020planck} and LSS data \cite{beutler2019primordial}, where the authors showed that existing LSS datasets like BOSS can surpass Planck in parts of the parameter space. However, the previous LSS analysis did not explore the entire parameter space that can be constrained with the available data. The maximum frequency range one can probe depends on the fundamental mode, $k_f$, which is the minimum wave number available in a survey: for a survey with arbitrary geometry and volume $V$, it can be approximated by $k_f=2\pi/V^{1/3}$. The importance of this scale arises because the measured power spectrum is convolved with the survey window function, which automatically damps any possible structure in $P(k)$ that is much finer than $k_f$. Here we analyze both the BOSS and eBOSS datasets using a binning in $k$ close to their fundamental mode, thus allowing us to scan for PF with the highest frequencies that can be probed given the limit set by the survey window. We will further elucidate this point in Section \ref{sec: window}. For linear features, we decided to use the parameter range specified in Table \ref{tab: PF_all_priors}. We emphasize that Eq.~(\ref{eqn: lin_feature}) is a phenomenological template: although being simple, its global oscillation in $k$ is theoretically unrealistic. Nevertheless, any relevant detection of Eq.~(\ref{eqn: lin_feature}) will ultimately indicate the presence of a sinusoidal running in the data. This conclusion will motivate the scan of other more realistic models with similar running, so Eq.~(\ref{eqn: lin_feature}) is definitely a robust starting point for primordial features analysis. We consider two of those realistic models in \S \ref{sub_sec: inflation_step} and \S \ref{subsec: varying_speed_sound}.
\label{sub_sec: linear_feature}

\subsection{Logarithmic features}
\begin{figure}[htbp]
    \centering
        \includegraphics[width = 0.48\textwidth]{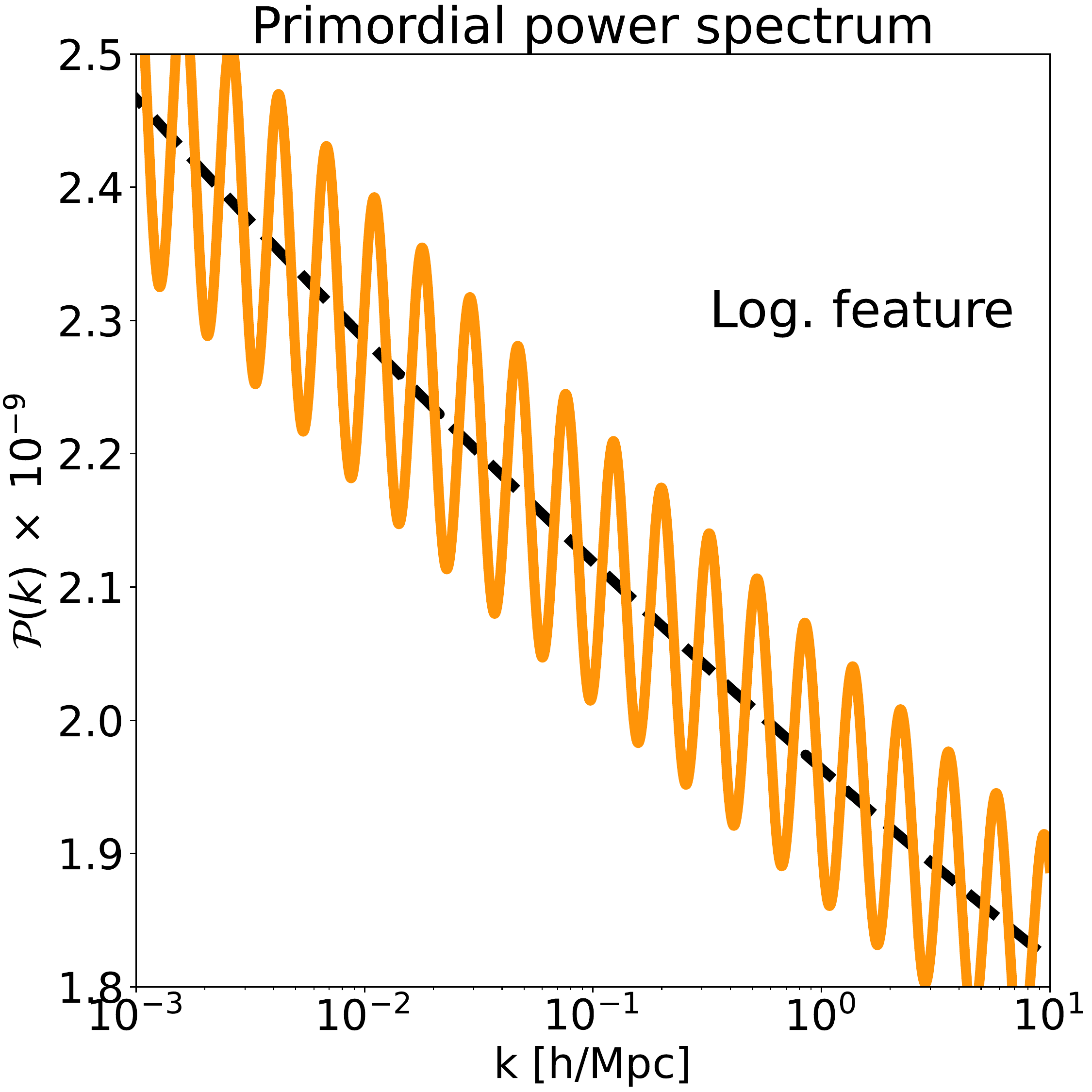}
    \includegraphics[width = 0.48\textwidth]{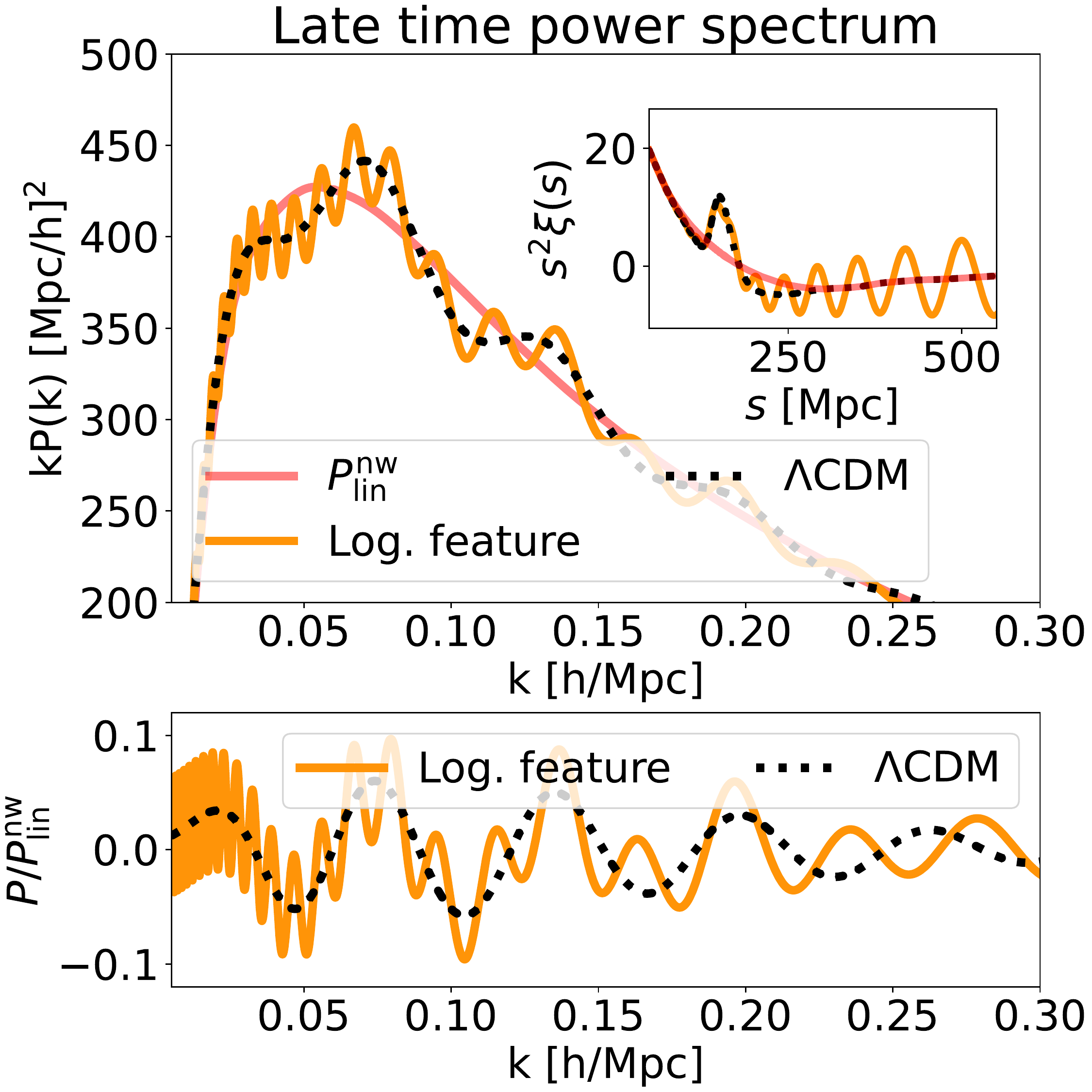}
    \caption{\textit{left:} The primordial curvature power spectrum in the vanilla model of inflation (black dashed line) and in the logarithmic feature case (solid orange line). For this plot, we used $A_{\rm log} = 0.05$, $\omega_{\rm log} = 35$, $\phi = 0$ and the BOSS high-$z$ window function; \textit{right:} An illustration of how the logarithmic feature shows up in the evolved galaxy field.}
    \label{fig: log_feature_plots}
\end{figure}
\label{sub_sec: log_features}
Axions are promising candidates for the inflaton. Contrary to the usual scalar field, axions have a continuous shift symmetry that protects their potential against radiative corrections, so they are safe from the eta-problem \cite{baumann2015inflation}, and the slow-roll conditions can be satisfied naturally. A known example of inflation being driven by axions is axion monodromy inflation \cite{chluba2015features,Flauger:2009ab,flauger2017drifting}. In that model, the axion potential is:
\begin{equation}
\label{eqn: axion_2}
    V(\phi) = \mu^3\phi - \Lambda^4\cos\left(\theta - \frac{\phi}{f}\right)\;,
\end{equation}
where $\mu$ is a mass scale. The presence of a sinusoidal modulation in Eq.\,(\ref{eqn: axion_2}) includes a new mechanism in the theory in which primordial non-Gaussianity can be generated: the oscillating perturbations can be in resonance with the oscillating inflaton background, which can lead to what is called \textit{Resonant non-Gaussianity} \cite{Chen:2008wn,Flauger:2010ja,Chen:2010bka, Chen:2006xjb}. This parametric resonance affects the primordial power spectrum, generating features \cite{chluba2015features, Flauger:2009ab, Flauger:2010ja}: 
\begin{equation}
    \delta \mathcal{P}_{\mathcal{R}}(k) = \kappa \cos \left[\frac{\phi_{*}}{f}-\frac{\ln \left(k / k_{*}\right)}{\widetilde{f}}\right]\;,
\end{equation}
where $\phi_{*}$ is the background value when $k_{*}$ exits the horizon, $\kappa = 3 \left( \Lambda^4/\mu^3f\right)(2\pi\tilde{f})^{1/2}$ and \begin{math}\smash{\tilde{f} = f\phi_{*}/M_{\rm Pl}^2} \end{math}. Given the theoretical importance of axions in both cosmology and particle physics, constraining these logarithmic signals is crucial. We scan for these features in galaxy clustering using the template
\begin{equation}
\label{eqn: log_features_1}
    \delta \mathcal{P}_{\mathcal{R}}(k) = A_{\log} \sin\left(\omega_{\log}\ln\left[\mathcal{K}\right] + \phi\right)\;,
\end{equation}
where  $\mathcal{K} = k/k_{\star}$ with the pivot scale $k_* = 0.05 \, \mathrm{Mpc}^{-1}$, and $A_{\rm log}$, $\omega_{\rm log}$ are the feature amplitude and frequency, respectively. In Figure \ref{fig: log_feature_plots}, we show how this primordial signal is present in both the primordial power spectrum, on the left panel, and in the late-time galaxy power spectrum, on the right panel. Notice that $\omega_{\rm log}$ has no units, so it can not be associated with a specific value in configuration space. 

Global logarithmic features have been widely tested with the CMB using different datasets. See \cite{Martin:2003sg, Martin:2004yi, Flauger:2009ab, 2012MNRAS.421..369M, Aich:2011qv, Peiris:2013opa} for an analysis using WMAP and \cite{akrami2020planck, Easther:2013kla} using Planck data. They have also been tested with LSS data \cite{Palma:2017wxu, beutler2019primordial, Ballardini:2022wzu}. In \cite{beutler2019primordial} it was shown that LSS can have stronger constraints than Planck in parts of the parameter space, which motivates an in-depth analysis of current and forthcoming data. Here we push the parameter space beyond what has been explored in \cite{beutler2019primordial} and to the maximum frequency that is accessible with current LSS datasets. A summary of the parameter space we scan is presented in Table \ref{tab: PF_all_priors}. When compared to previous LSS analysis \cite{beutler2019primordial} we have two differences. Firstly, as briefly scrutinized at the end of \S\ref{sub_sec: linear_feature}, our power-spectrum binning allows us to scan for features over a larger frequency range. Secondly, we decided to adopt the phase parametrisation instead of two amplitudes because it is more natural to impose a non-informative prior (see Appendix \ref{app: phase_impact} for more details).

\subsection{Inflation with a step}
\label{sub_sec: inflation_step}

Oscillatory features can also be generated when the inflaton potential passes through a step in the potential \cite{adams2001inflationary}. Here we use the results of \cite{miranda2014inflationary} to search for these features in large-scale data and refer the reader to this paper for more details. Hereafter this model will be referred to as \textit{step features}.
\begin{figure}[htbp]
    \centering
        \includegraphics[width = 0.48\textwidth]{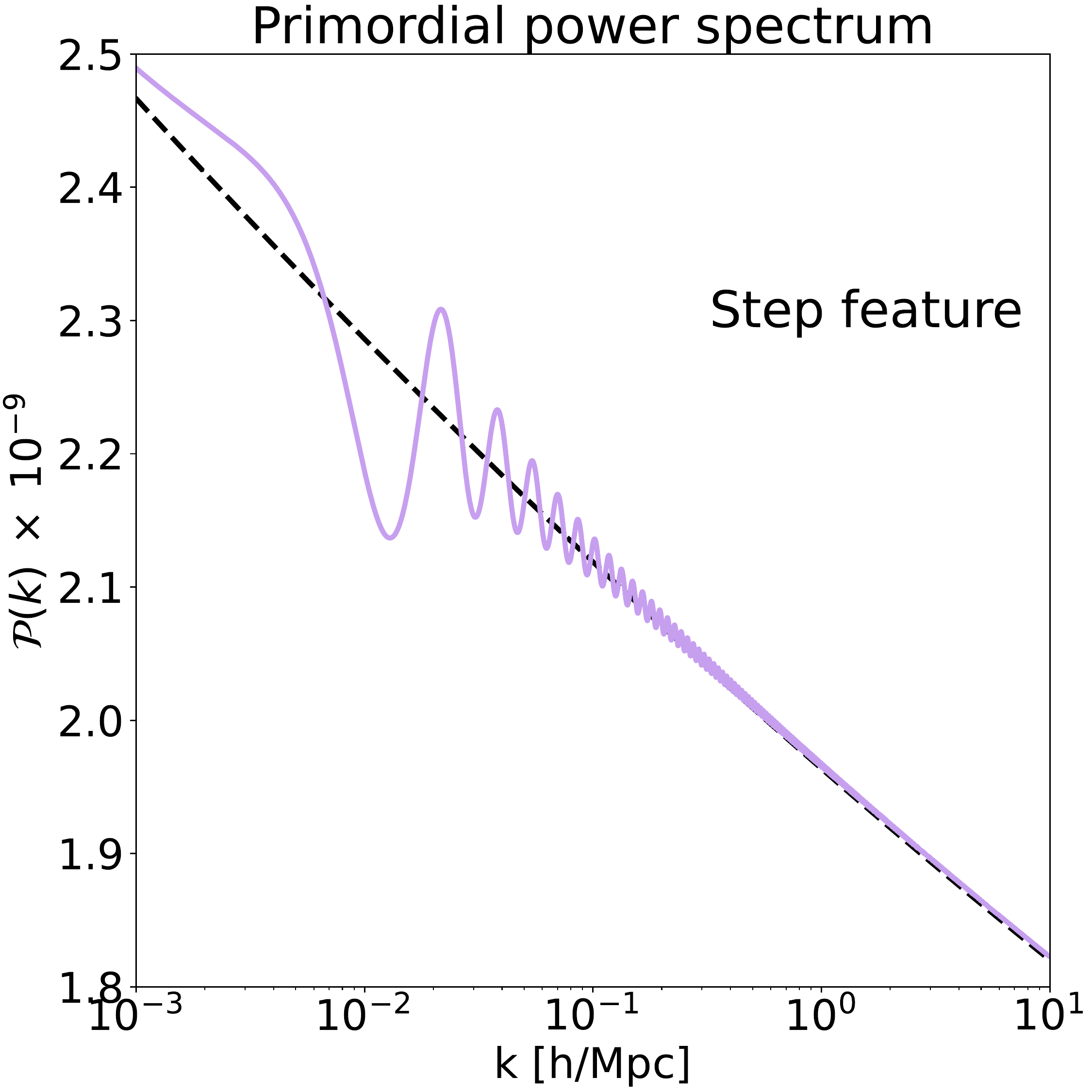}
    \includegraphics[width = 0.48\textwidth]{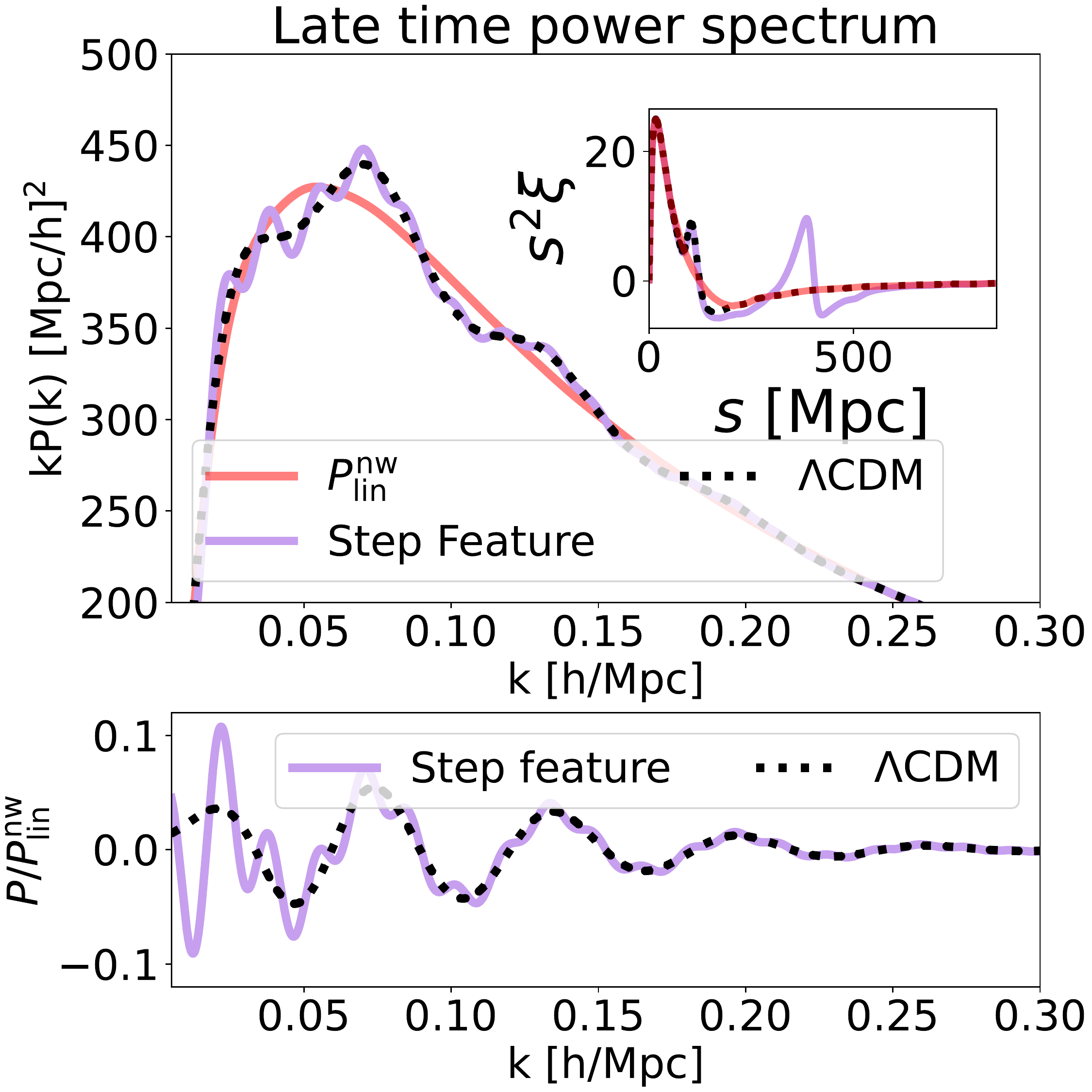}
    \caption{\textit{left:} The primordial curvature power spectrum in the vanilla model of inflation (black dashed line) and the predictions for a step in the potential case (solid purple line). For the step feature, we used $\mathcal{A}_s = 0.05$, $\omega_s = 200\;\mathrm{Mpc}$ and $x_s = 100$; \textit{right:} An illustration of how this feature shows up in the evolved galaxy field.}
    \label{fig: step_potential_plots}
\end{figure}

This feature signal was tested in previous work with CMB \cite{akrami2020planck} data. Although predictions for this model were presented in \cite{palma2018constraints}, it has never previously been tested with LSS datasets. We perform a full MCMC analysis using the BOSS and eBOSS datasets and test the pipeline with mock catalogues. Moreover, we also consider the correct effect of the survey window function. As we will discuss in Section \ref{sec: window}, the window function convolution imposes a strong limitation on our constraints when scanning for highly oscillatory features. For this model, the impact is even more pronounced since the feature amplitude also decreases for higher frequencies. 

An example modification of the galaxy correlation function by this model and its associated primordial power spectrum is shown in Figure \ref{fig: step_potential_plots}. It is analytically described by
\begin{eqnarray}
\label{eqn: step_potential_version_1}
\ln \delta\mathcal{P}_{\mathcal{R}}^{\mathrm{s}}(k) =\ln \mathcal{P}_{\mathcal{R}}^{0}(k)+\mathcal{I}_{0}(k)+\ln \left(1+\mathcal{I}_{1}^{2}(k)\right)\;,
\end{eqnarray}
where the kernels are given by:
\begin{eqnarray}
\mathcal{I}_{0}&=&\mathcal{A}_{\mathrm{s}} \mathcal{W}_{0}\left(k \omega_s\right) \mathcal{D}\left(\frac{k\omega_s}{x_{\mathrm{s}}}\right), \\
\mathcal{I}_{1}&=&\frac{1}{\sqrt{2}}\left[\frac{\pi}{2}\left(1-n_{\mathrm{s}}\right)+\mathcal{A}_{\mathrm{s}} \mathcal{W}_{1}\left(k \omega_s\right) \mathcal{D}\left(\frac{k \omega_s}{x_{\mathrm{s}}}\right)\right]\;.
\end{eqnarray}
The oscillatory terms $\mathcal{W}_0$ and $\mathcal{W}_1$ are
\begin{eqnarray}
\label{eqn: step_w0}
\mathcal{W}_{0}(x)&=&\frac{1}{2 x^{4}}\left[\left(18 x-6 x^{3}\right) \cos 2 x+\left(15 x^{2}-9\right) \sin 2 x\right]\;;\\
\label{eqn: step_w1}
\mathcal{W}_{1}(x)&=&-\frac{3}{x^{4}}(x \cos x-\sin x)\left[3 x \cos x+\left(2 x^{2}-3\right) \sin x\right]\;;
\end{eqnarray}
and $\mathcal{D}(x)$ is a damping term that controls the localization of the feature:
\begin{equation}
\mathcal{D}(x)=\frac{x}{\sinh x}\;.
\end{equation}
For more details about this model we refer the reader to \cite{miranda2014inflationary}. To put these corrections in the correct format for our pipeline we will rewrite them as
\begin{equation}
\mathcal{P}_{\mathcal{R}}^{\mathrm{s}}(k) = \mathcal{P}_{\mathcal{R}}^{0}(k)\left[1+\mathcal{I}_{1}^{2}(k)\right]e^{\mathcal{I}_{0}(k)}\;.
\end{equation}
Since $\mathcal{I}_0$ is an oscillation around zero its exponential will be an oscillation around unity, so we rewrite the model as
\begin{equation}
    e^{\mathcal{I}_0(k)} \equiv 1 + \delta \mathcal{I}_0\;,
\end{equation}
so the equation to be tested becomes
\begin{equation}
\mathcal{P}_{\mathcal{R}}^{\mathrm{s}}(k) = \mathcal{P}_{\mathcal{R}}^{0}(k)\left[1+\mathcal{I}_{1}^{2}(k) + \delta \mathcal{I}_0 +  \mathcal{I}_{1}^{2}(k) \delta \mathcal{I}_0\right] \equiv \mathcal{P}_{\mathcal{R}}^{0}(k)\left[ 1 + \delta\mathcal{P}\right]\;.
\end{equation}
The priors we use for this model are presented in Table $\ref{tab: PF_all_priors}$.
\begin{figure}[t!]
    \centering
        \includegraphics[width = 0.48\textwidth]{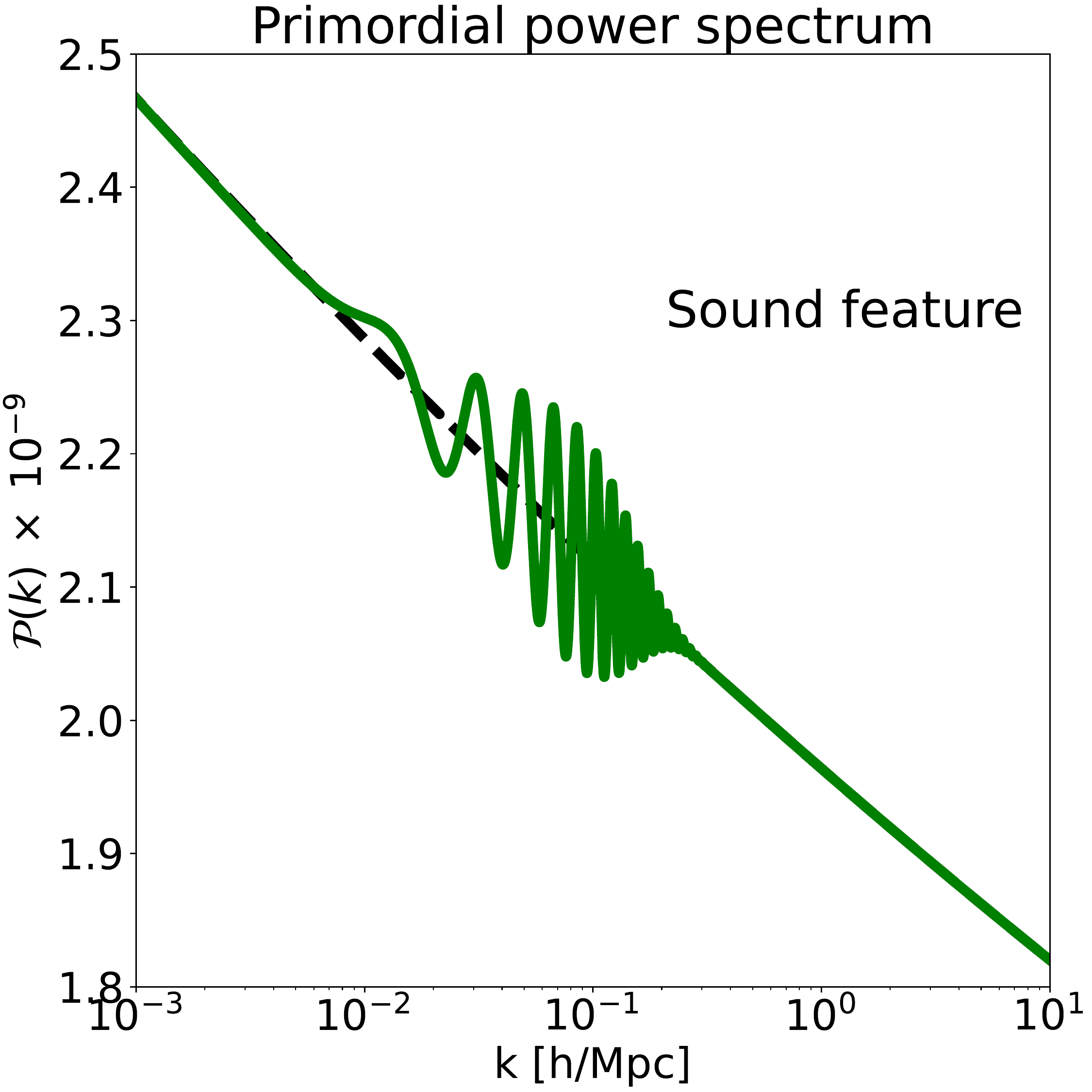}
    \includegraphics[width = 0.48\textwidth]{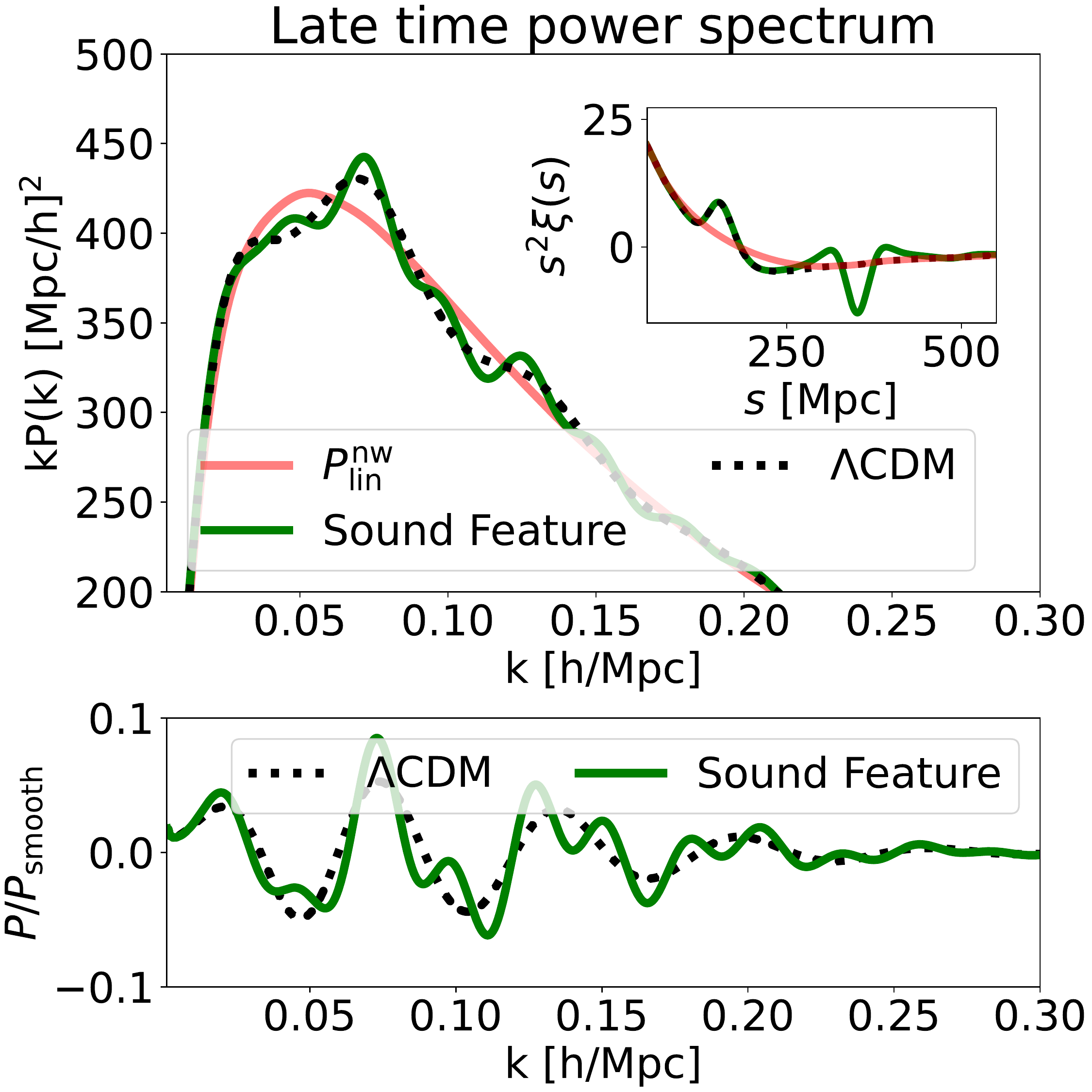}
    \caption{\textit{left:} The primordial curvature power spectrum in the vanilla model of inflation (black dashed line) and in the sound feature case (solid pink line). For this plot we used $\mathcal{B} = -0.5$, $|\tau_f| = 175\,\mathrm{Mpc}$ and $\ln \beta = 6$; \textit{right:} An illustration of how the sound feature shows up in the evolved galaxy field.}
    \label{fig: sound_feature_plots}
\end{figure}
\subsection{Variation in the speed of sound with a Gaussian profile}
\label{subsec: varying_speed_sound}
In an effective field theory framework, the inflaton field is necessarily coupled to heavier fields. These fields can be integrated out, leading to an effective single-field model \cite{Burgess:2012dz, Achucarro:2012sm, Achucarro:2012yr}. In this effective field theory, derivative couplings can lead to a sharp reduction in the speed of sound. These variations generate features in both the primordial power spectrum and bispectrum of the adiabatic modes \cite{Gong:2014spa, Cai:2013gma, Bartolo:2013exa}.
In \cite{Achucarro:2010jv, Achucarro:2014msa,  Achucarro:2013cva} the authors computed the consequences for the primordial power spectrum of a time variation in the sound speed. For a departure of the speed of sound from unity they concluded that the primordial power spectrum is
\begin{equation}
\label{eqn: sound_spectrum}
\begin{gathered}
\frac{\Delta \mathcal{P}_{\mathcal{R}}}{\mathcal{P}_{\mathcal{R}, 0}}=\frac{1}{36}\left[\sin \left(2k  \tau_f\right)+\frac{1}{k\tau_f} \cos \left(2k  \tau_f\right)\right] \mathcal{D}_S
-\frac{1}{72}\left[\frac{1}{\left(k  \tau_f\right)^2} \sin \left(2k  \tau_f \right)\right] k \frac{d}{d k} \mathcal{D}_S
\end{gathered}\;,
\end{equation}
with the envelope $\mathcal{D}_{S}$ being
\begin{equation}
\mathcal{D}_S=\frac{4 \sqrt{\pi} \mathcal{B} k \tau_{f}}{\sqrt{\beta}} \exp \left(-\frac{k^2 \tau_{f}^2}{\beta}\right)\;.
\end{equation}
The parameters $\mathcal{B} < 0$ represents the feature amplitude, $\tau_f$ is the characteristic time of the feature, which sets its frequency and the parameter $\beta > 0$ controls the sharpness of the feature. This PF model has been tested by \cite{Achucarro:2013cva} and \cite{Hu:2014hra, Palma:2017wxu} using CMB and LSS data, respectively. In total, it has three free parameters: $\mathcal{B}$, $\tau_f$, and $\beta$, which makes this model difficult to analyze. For this reason, we follow the discussion in \cite{Achucarro:2014msa} to motivate a preferred region in parameter space to look for this feature. In \cite{Cannone:2014qna} it is argued that for the perturbative treatment of the effective single-field dynamics to remain valid, the sharpness parameter should satisfy $\ln \beta < 14$ \cite{Achucarro:2014msa}. We then decided to scan over the region $2.2 < \ln \beta < 10$. We summarize the priors in Table \ref{tab: PF_all_priors}.

\section{Datasets}
\label{sec: data_used}

In this work, we scan for signals of primordial features in large-scale structure. Our results are obtained from a joint analysis of both the BOSS low-z and high-z datasets as well as the  eBOSS QSO sample. In this section, we briefly describe the aspects of each dataset, which are summarized in Table \ref{tab: products}. \footnote{All the products used in this work can be downloaded at \url{https://fbeutler.github.io/hub/hub.html}}
\begin{figure}
    \centering
    \includegraphics[width = 0.48\textwidth]{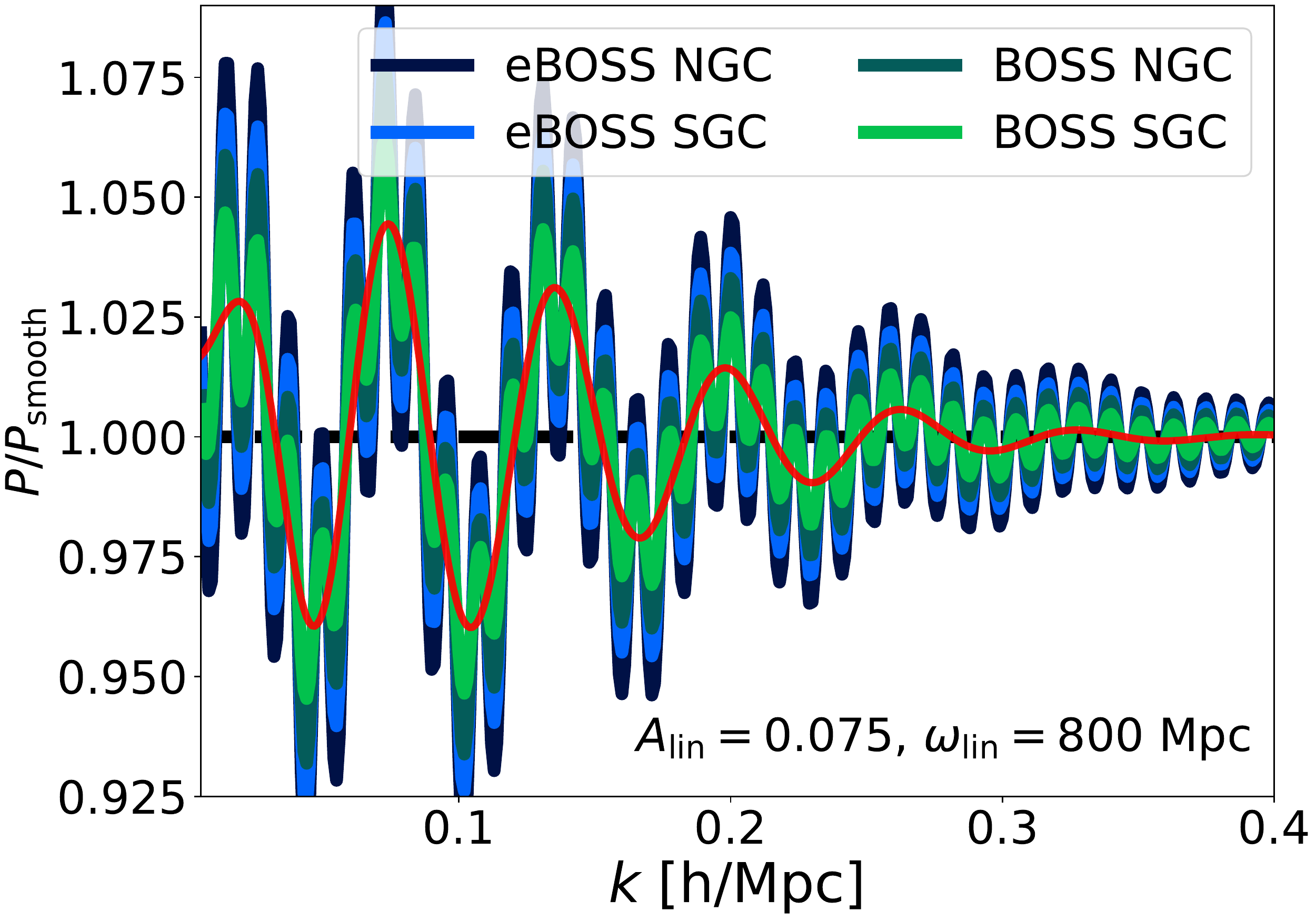}
    \includegraphics[width = 0.48\textwidth]{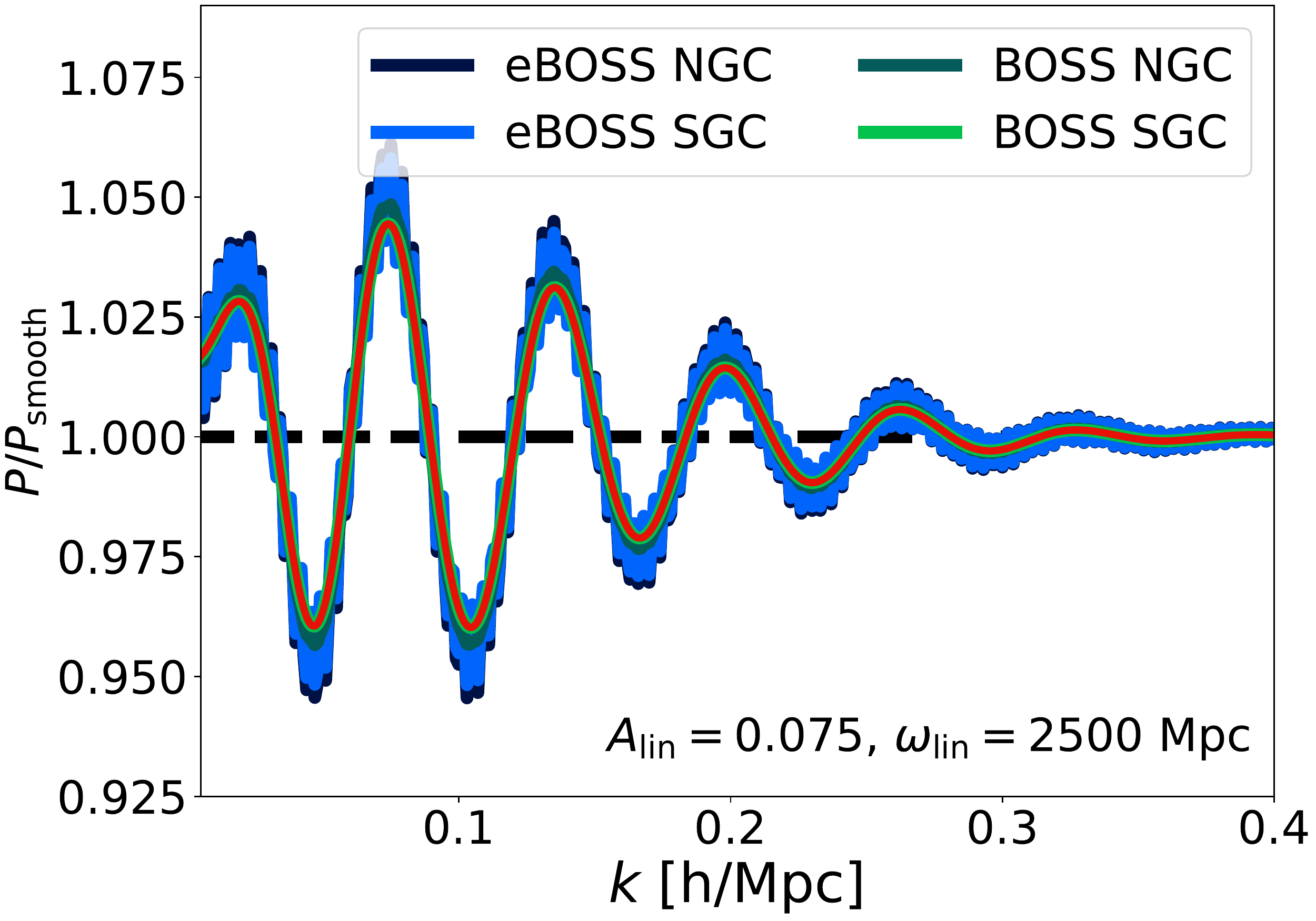}
    \includegraphics[width = 0.48\textwidth]{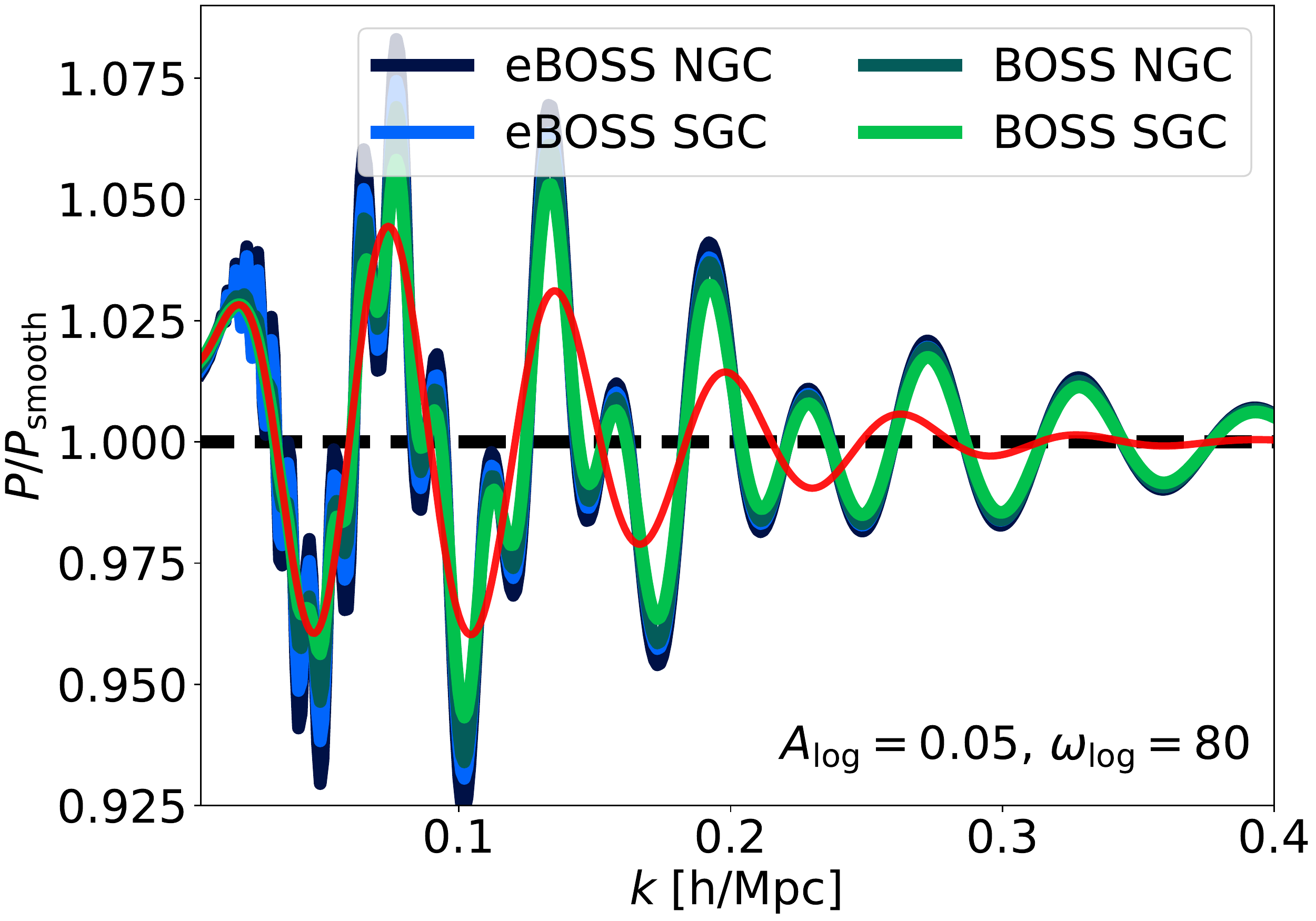}
    \includegraphics[width = 0.48\textwidth]{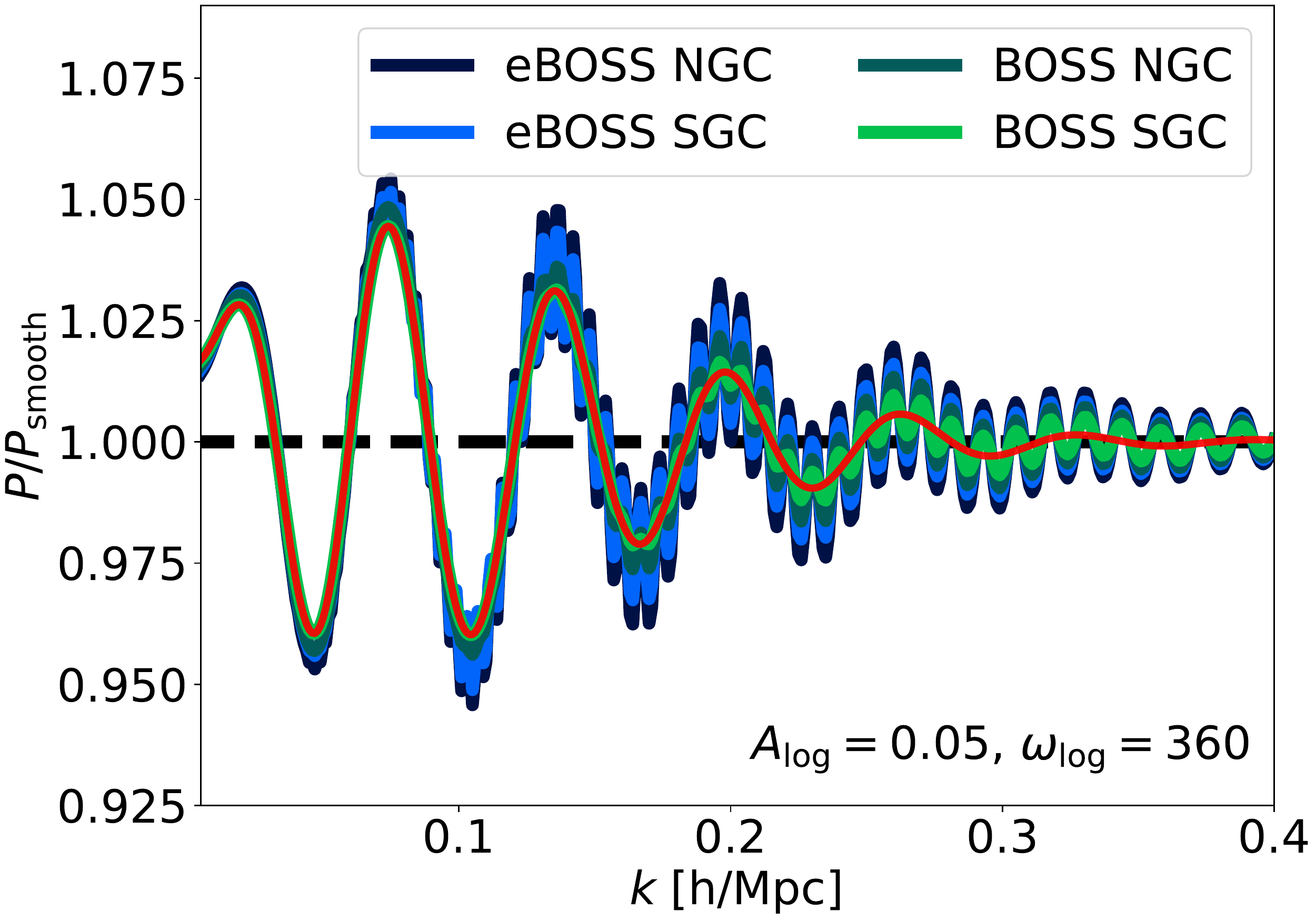}
    \caption{We show how the survey window function affects primordial oscillatory features in large-scale structure data. We generated fictitious linear and logarithmic oscillations (fixed amplitude but different frequencies) and convolved them with the window functions we use in this work (blue and green lines for eBOSS and BOSS, respectively). What we plot above is the wiggly part of the resulting power spectrum. The red line is the spectrum with no primordial features, so only the usual BAO oscillations are present. Note that overall the window function smearing is more important for oscillations with large frequencies. For linear features, the BOSS window function completely erases the primordial feature signal for $\omega_{\rm lin } = 2500\;\mathrm{Mpc}$. For logarithmic features, the window function effect is not straightforward, since the frequency is a function of $k$. Nevertheless, the smearing always occurs from small values of $k$. Regardless of the models considered in this work, all PF signals are erased in the limit $\omega \rightarrow \infty$.}
    \label{fig: window_function_impact}
\end{figure}
\begin{table}[]
\centering
\begin{tabular}{@{}ccccc@{}}
\toprule
                         & \textbf{Lin. feature} & \textbf{Log. feature} & \textbf{Step feature} & \textbf{Sound feature} \\ \midrule
$A_{\rm X}$              & {[}$-1$, 1{]}           & {[}$-1$, 1{]}           & ...                   & ...                    \\
$\omega_{\rm X}$         & {[}100, 4000{]}       & {[}10, 360{]}         & ...                   & ...                    \\
$\phi_{\mathrm {X}}/\pi$ & {[}$-0.5$, $0.5${]}   & {[}$-0.5$, $0.5${]}   & ...                   & ...                    \\ \midrule
$A_{\rm step}$           & ...                   & ...                   & {[}$-1$, 1{]}           & ...                    \\
$\omega_{\rm step}$      & ...                   & ...                   & {[}100, 1000{]}       & ...                    \\
$x_s$                    & ...                   & ...                   & {[}10, 650{]}         & ...                    \\ \midrule
$\mathcal{B}$             & ...                   & ...                   & ...                   & {[}$-1$, 0{]}          \\
$\ln \beta$              & ...                   & ...                   & ...                   & {[}2.2, 10{]}          \\
$|\tau_f|$               & ...                   & ...                   & ...                   & {[}100, 900{]}         \\ \bottomrule
\end{tabular}
\caption{Ranges for the uniform priors we use for the four primordial feature models we investigate in this paper. The parameters $\omega_{\rm lin}$, $\omega_{\rm step}$, $\tau_f$ and $x_s$ are in units of Mpc.}
\label{tab: PF_all_priors}
\end{table}
\subsection{BOSS data}
The Baryon Oscillation Spectroscopic Survey (BOSS) was part of the Sloan Digital Sky Survey III (SDSS-III) and, at the present moment, is the dataset with the largest effective volume \cite{dawson2012baryon, eisenstein2011sdss, smee2013multi, SDSS:2006srq}. Here, we use data release 12 (DR12) \cite{Reid:2015gra}, which consists of 1\,198\,006 galaxies divided into two patches in the sky, the North Galactic Cap (NGC) and the South Galactic Cap (SGC). We follow the standard approach and split the data in two (independent) redshift bins: the low-$z$ sample, with $0.2 < z < 0.5$, and the high-$z$ sample, with $0.5 < z < 0.75$. As we will explain in \S\ref{sec: window}, the $k$-binning choice is relevant for the analysis of the primordial features, since smaller bins usually allow us to probe PF with higher frequencies. We can not decrease the bin size indefinitely, since it is limited by convolution with the window function: this has a width of order the survey fundamental mode, $k_f$, and so modes separated by less than this amount will be strongly correlated. Moreover, the estimation of the covariance matrix will require more mock catalogues in order to be reliable. With that in mind, we decided to use a binning that is close to the survey fundamental mode defined in \cite{Beutler:2021eqq}, which is $\Delta k = 0.002 h \,\mathrm{Mpc}^{-1}$ for both NGC and SGC~\footnote{In principle one could use a different binning for NGC and SGC since they have different volumes, which allows us to probe primordial features with higher frequencies. But the signal at such high frequencies is already dominated by the eBOSS sample so this would not greatly affect the final result.}. Since the inflationary signal is isotropic, we only consider the power spectrum monopole in our analysis. In recent work \cite{Ballardini:2022wzu}, the authors included the quadrupole in the analysis. Although the feature signal is isotropic, the window function can leak some information from the monopole into the higher-order multipoles. However, this leakage is small and the quadrupole naturally has a smaller signal to noise than the monopole. Moreover, the feature amplitude is expected to be small, so the inclusion of the quadrupole will not bring significant additional information. We also emphasize that an anisotropic analysis requires only a trivial extension of the pipeline, which can easily be implemented in the future.  

Another product we will use is the MultiDark-Patchy mock catalogues that reproduce the BOSS DR12 dataset \cite{Kitaura:2015uqa}. These catalogues are obtained using approximate solvers for the complicated gravitational collapse models, reproducing a similar galaxy number density, survey geometry, and selection function. We refer the reader to the reference above for more details. Their use here is twofold. Firstly, we use the $997$ NGC and $1000$ SGC mock catalogues to derive the covariance matrix required for the MCMC analysis. Secondly, since these mocks were generated without primordial features, they can be used as a validation test of our analysis pipeline for the case without features.

To further improve the sensitivity to oscillatory signals in the BOSS data, we perform a field reconstruction as described in \cite{Eisenstein:2006nk}. This helps reduce the effects of non-linear displacements, which as described in \S\ref{sec: non_linear_damping}
can smear out oscillations present in the power spectrum. In addition, the displacements can also cause a change in the feature scale, shifting the oscillation frequency. In \cite{Li:2021jvz}, the authors tested the impact of a similar field reconstruction scheme and concluded that it alleviates these two problems.
\begin{figure}[t!]
    \centering
    \includegraphics[width = 0.5\textwidth]{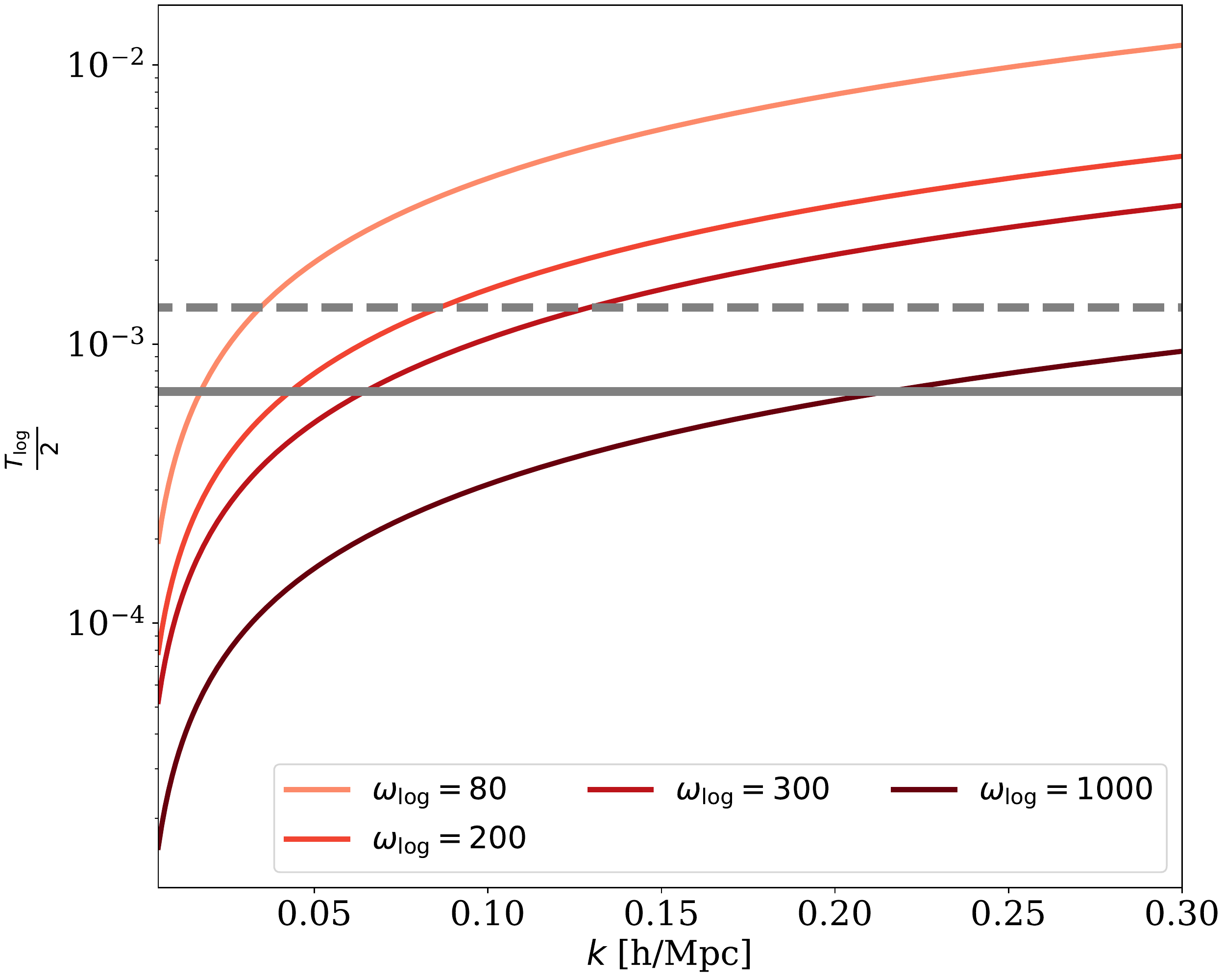}
    \caption{The oscillation half period for logarithmic features. Notice that the frequency of the oscillation varies with $k$. The solid and dashed grey lines represent the half period of linear oscillations with frequencies equal to the Nyquist frequency for $\Delta k = 0.002 h\, \mathrm{Mpc}^{-1}$ and $\Delta k = 0.001 h \, \mathrm{Mpc}^{-1}$, respectively.}
    \label{fig:nyquist_log}
\end{figure}
\subsection{eBOSS data}
\begin{table}[]
\centering
\begin{tabular}{|l|c|c|c|c|l|}
\hline
Sample            & \multicolumn{1}{l|}{$[z_{\rm min}, z_{\rm max}]$} & \multicolumn{1}{l|}{$z_{\rm eff}$} & \multicolumn{1}{l|}{$\Delta k \; [h\mathrm{Mpc}^{-1}]$} & \multicolumn{1}{l|}{\begin{tabular}[c]{@{}l@{}}$[k_{\rm min}, k_{\rm max}]$ \\ $ \; [h \mathrm{Mpc}^{-1}]$\end{tabular}} & \begin{tabular}[c]{@{}l@{}}Nyquist Freq. \\ $\; [\mathrm{Mpc}]$\end{tabular} \\ \hline
BOSS low-$z$       & {[}0.2, 0.5{]}                                    & 0.38                               & 0.002                                                   & {[}0.01, 0.3{]}                                                                                                          & \quad 2323                                                                   \\ \hline
BOSS high-$z$       & {[}0.5, 0.75{]}                                   & 0.61                               & 0.002                                                   & {[}0.01, 0.3{]}                                                                                                          & \quad 2323                                                                  \\ \hline
eBOSS Quasars & {[}0.8, 2.2{]}                                    & 1.52                               & 0.001                                                   & {[}0.02, 0.23{]}                                                                                                         & \quad 4647                                                                                                   \\ \hline
\end{tabular}
\caption{Summary of the data used in this work, presented in Section \ref{sec: data_used}. For a binned data vector there is a maximum frequency one can probe. This fundamental limit due to the discrete nature of the data is called the Nyquist frequency, defined as $\omega_{\rm Ny} = \pi/\Delta k$. We adopt the same fiducial cosmology used in the BOSS and eBOSS official analysis. To convert between units with and without the Hubble factor, we use $h_{\rm fid} = 0.676$, which is common to both fiducial cosmologies. In Figure \ref{fig: lin_feature_result} we also show BOSS constraints beyond the Nyquist frequency defined above. In order to do this, we switch the BOSS binning to be the same as eBOSS for frequencies above $\omega_{\rm lin} \approx 2323\, \mathrm{Mpc}$.}
\label{tab: products}
\end{table}

The extended Baryon Oscillation Spectroscopic Survey (eBOSS) \cite{Dawson:2015wdb, Ross:2020lqz} is part of SDSS-IV \cite{blanton2017sloan}. Here we use the QSO sample, which is composed of $\sim 350\,000$ quasars. Although the number density is considerably lower when compared to the BOSS dataset described previously, it covers a larger cosmic volume. The survey volume is crucial when looking for primordial features, and affects the analysis through the window function. A larger cosmic volume allows us to probe features with higher frequencies/smaller wavelength. The eBOSS QSO dataset is also divided between two galactic caps (NGC and SGC) and has no redshift overlap with the BOSS galaxy sample. We also use QSO mock catalogues to validate our pipeline for that dataset \cite{Zhao:2020bib}. These mocks were obtained by implementing the Extended Zel’dovich (EZ) approximate N-body simulation technique \cite{Chuang:2014vfa}. The eBOSS QSO sample has a cosmic volume that is considerably larger than the
BOSS sample, which allows a smaller $k$-binning: $\Delta k = 0.001 h \,\mathrm{Mpc}^{-1}$. We use 1000 mock catalogues to compute the eBOSS QSO covariance matrix.

\subsection{Window function}
\label{sec: window}
All data collected from surveys are formed as a combination of the true density fluctuations multiplied by weights that account for the survey geometry and systematic effects. In Fourier space this weight appears as a convolution, modifying the shape of the observed power spectrum. 
We estimate the window function from the random catalogue and convolve the theory with it, following the steps described in \cite{Wilson:2015lup}. In summary, the steps are:
\begin{itemize}
    \item Compute the theory in Fourier space and Hankel transform it to obtain the theoretical correlation function.
    \item Calculate the convolved theoretical correlation function by multiplying it with the window function. For the monopole, this multiplication is trivial because we are ignoring the quadrupole.
    \item Perform an inverse Hankel transform on the outcome of the last step  to obtain the convolved power spectrum. This is the quantity we use to compare with the data.
\end{itemize}
In Figure \ref{fig: window_function_impact} we show a series of examples of the influence of the window function for both linear (first four panels on the top) and logarithmic features (last four panels). The greyish blue line is the usual BAO wiggle, and the other lines represent the results with additional primordial features with fixed amplitude $A_{\rm lin, log} = 0.075$. In general, the window function smears out any oscillatory signal, and this smearing is stronger for higher-frequency oscillations. This trend can be seen for both linear and logarithmic oscillations. For instance, for linear frequencies $\omega_{\rm lin} \approx 800\,\mathrm{Mpc}$ the resulting oscillatory pattern is almost the same for all survey geometries, with the damping being stronger for BOSS than eBOSS. As the frequency is increased, the BOSS window function tends to degrade the oscillations with respect to eBOSS even further, and the signal is completely removed for $\omega_{\rm lin} \approx 2500 \, \mathrm{Mpc}$. This limiting frequency is higher for the eBOSS QSO dataset because it has a greater cosmic volume. For small frequency values, the BOSS constraints will overcome the eBOSS ones, simply because BOSS has a greater signal-to-noise; but because  BOSS loses sensitivity to high frequencies first, we expect eBOSS constraints to dominate at high frequencies. For the logarithmic features, a similar trend is present.  In Figure \ref{fig:nyquist_log} we plot the separation between zeroes for a logarithmic oscillation with different frequencies. The horizontal lines represent the Nyquist frequencies, $\omega_{\rm Ny} = \pi/\Delta k$, for $\Delta k = 0.002h\,\mathrm{Mpc}^{-1}$ and $0.001 h\, \mathrm{Mpc}^{-1}$. A survey window function with a fundamental mode close to these values will erase all signal whose separation of zeros is below $\omega_{\rm Ny}$. Therefore, according to Figure \ref{fig:nyquist_log}, the smearing due to the window function for logarithmic oscillations will occur from lower values of $k$. This is confirmed in Figure \ref{fig: window_function_impact}, and as the frequency increases the signal starts to disappear first for small values of $k$.

\section{Statistical methods and analysis pipeline}
\label{sec: methods}

In this section, we explain how we test the primordial feature models and compare them with the reference single-field slow-roll model. 
\begin{table}[]
\centering
\begin{tabular}{c|c|c|c}
                  & BOSS high-$z$         & BOSS low-$z$                      & eBOSS QSO         \\ \hline
$\alpha_{\rm iso}$ & 0.9887 $\pm$ 0.0089 & 0.997 $\pm$ 0.0105 & 1.023 $\pm$ 0.020 \\
Official results & 0.9887 $\pm$ 0.0087 & 1.000 $\pm$ 0.010
& 1.025 $\pm$ 0.020
\end{tabular}
\caption{Results for a preliminary BAO isotropic analysis using our pipeline with no primordial features. The measurements of $\alpha_{\rm iso}$ are in very good agreement with the ones quoted by previous works \cite{BOSS:2016hvq, Neveux:2020voa}}
\label{tab: BAO-only}
\end{table}

\subsection{Performance metrics}
\label{subsec: stat_methods}

If the hypothesis of primordial features is correct, then there will be two kinds of oscillations in the power spectrum: the first one is the usual baryon acoustic oscillation (BAO) feature, whose physics is well-known and which is one of the main science drivers for modern galaxy redshift survey experiments. The second one is the primordial feature, which couples to the BAO and can be written as in Eq.\,(\ref{eqn: final_template}).  We take as reference model the simple situation in which only the BAO oscillation exists; we call this model $\mathcal{M}_0$. The main goal is then to answer the following question: \textit{Given some primordial feature model, $\mathcal{M}_1$, from the options set out in Section \ref{sec: primordial_features} and the best existing clustering data, how likely is $\mathcal{M}_1$ to be true relative to $\mathcal{M}_0$?}. This is a model comparison problem, which can be solved using standard Bayesian methods \cite{Trotta:2017wnx, Trotta:2005ar, Jenkins:2011va}. If $\mathcal{M}_0$ and $\mathcal{M}_1$ are taken as equally likely a priori, then the relative probability (odds ratio) between the two models is given by the Bayes factor, $B_{01}$:
\begin{equation}
    B_{01} = \frac{p(D|\mathcal{M}_0)}{p(D|\mathcal{M}_1)}\;,
    \label{eqn: bayes_factor_definition}
    \end{equation}
where $p(D|\mathcal{M}_i)$ is the evidence of model $i$, given some data $D$. Usually, the computation of the evidence is numerically costly, since it involves the evaluation of a multi-dimensional integral. Nevertheless, note that the primordial feature models are extensions of $\mathcal{M}_0$, and they coincide when the feature amplitude, $A$, vanishes. In other words $\mathcal{M}_0$ is nested inside $\mathcal{M}_1$, and it has been shown that Eq.\,(\ref{eqn: bayes_factor_definition}) can then be evaluated using the Savage-Dickey ratio \cite{Dickey1971TheWL,Trotta:2005ar}:
\begin{equation}
\label{eqn: B01_Savage_Dickey}
    B_{01} = \frac{p(A|D, \mathcal{M}_1)}{p(A|\mathcal{M}_1)}\Big|_{A = 0}\;,
\end{equation}
where $p(A|D,\mathcal{M}_1)$ is the marginal posterior for $A$ under model $\mathcal{M}_1$ given the data and $p(A|\mathcal{M}_1)$ is the prior for $A$ in that model. Hence the only ingredient we need in order to compute the odds ratio is the marginal posterior of the feature amplitude for each model. We estimate the posterior using samples obtained via the \texttt{emcee} package \cite{emcee}. For each analysis, we run four parallel chains and use the Gelman-Rubin criterion for convergence. After the chains converge, we apply a 40\% burn-in and combine them into one final chain. Further analysis is performed using the package \texttt{getdist} \cite{Lewis:2019xzd}. For step and sound features, the amplitude is positive-definite, so the samples we obtain near $A = 0$ might not be representative.  For these models, we therefore computed the Bayes factor using the \texttt{pocoMC} software \cite{Karamanis:2022alw, Karamanis:2022ksp}. When necessary, we will also quote the minimum $\chi^2$ for different models. This value is found using the \texttt{iminuit} software \cite{iminuit}, which is a Python code based on the C$++$ library \texttt{Minuit} \cite{James:1975dr}. Our main results are obtained by the joint constraints from all three data products we use: BOSS low-$z$, high-$z$ and the eBOSS QSO sample, which we assume to be uncorrelated, so the final combined posteriors are obtained by multiplying each individual posterior. 

In addition to the Bayes factor, we also compute the 95\% credible intervals for the primordial feature amplitude as a function of its frequency. To do this, we first split the MCMC samples into frequency bins to obtain the binned posterior. Hence, we analyze the marginal posterior on the feature amplitude. We start integrating it around $|A| = 0$ and stop when the cumulative probability reaches $95 \%$. Thus, for a given frequency bin $i$, we will have a value $A_i$ such that $p(|A|\leq A_i|D) = 95\%$. High values of $A_i$ indicate the marginal posterior on $A$ for the frequency bin $i$ is distant from the origin, which can be associated with a possible detection. 


\begin{figure}
    \centering
    \includegraphics[width = \textwidth]{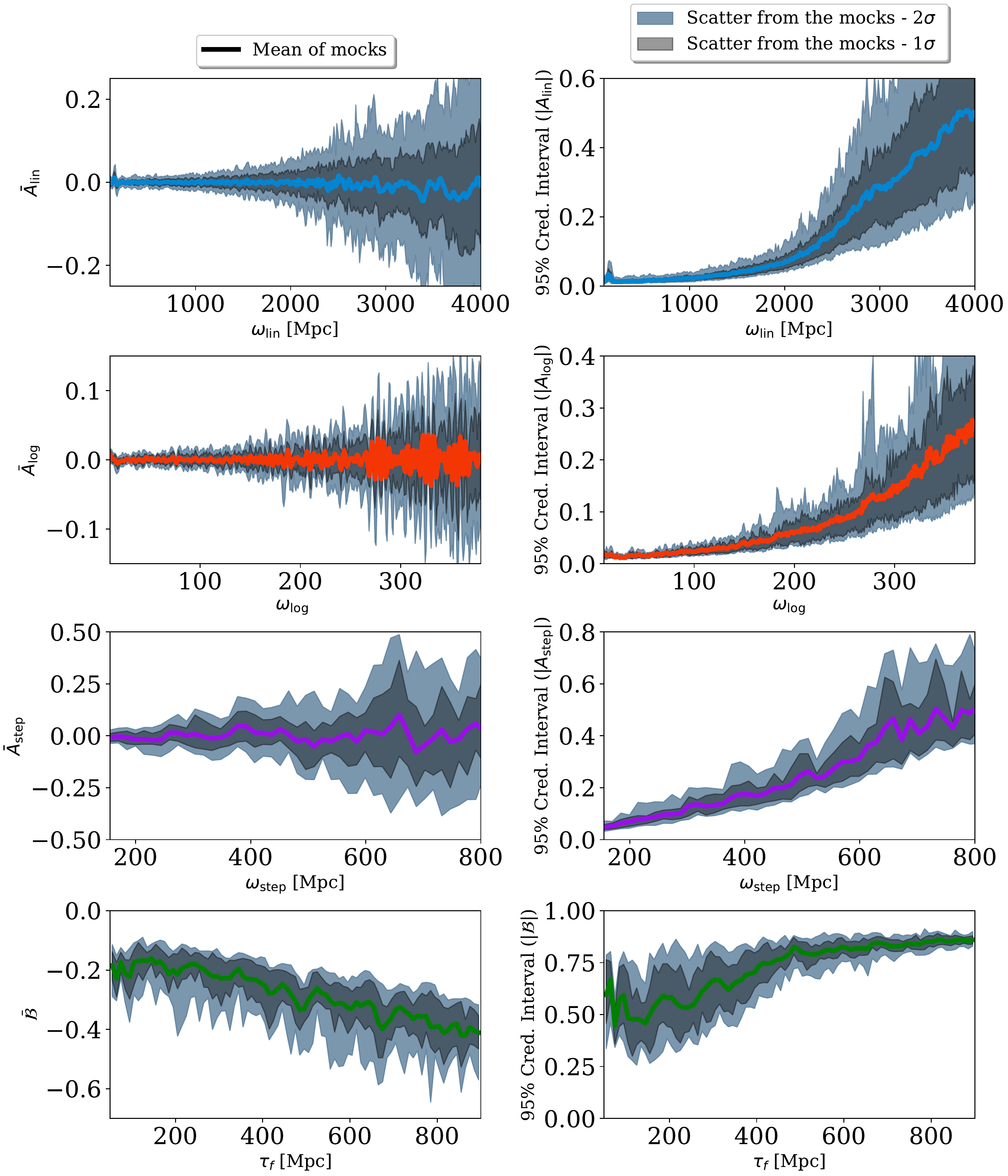}
    \caption{The combined result of the analysis when looking for linear, logarithmic, step and sound features using mock catalogues of BOSS high-$z$, low-$z$, and eBOSS QSO sample. In the left panels, we show the median of the posterior for each mock individually (light colours), as well as their sample mean (dark colour). In the right panels, we show the 95\% credible interval for the PF amplitude, $|A|$, obtained by integrating the posterior.}
    \label{fig: test_mocks}
\end{figure}

\subsection{Validating the pipeline}
The mocks are constructed so that the only feature on top of the smooth broadband power spectrum is the BAO signal. However, apparent primordial features can be generated by noise, and it is convenient to use an ensemble of mocks to  estimate the probability of observing a given signal just by chance, particularly given the need to allow for the look-elsewhere effect as discussed below in \S\ref{subsec: look-elsewhere}. We perform such a test for all models that we introduced in Section \ref{sec: primordial_features} and discuss the results here. The comparison of the mock analysis with data is performed in Section \ref{sec:results}.

Starting with linear features, we scan the range $\omega_{\rm lin} \in [100, 4000]\; \mathrm{Mpc}$ and estimate a posterior on $A_{\rm lin}$ using the MCMC samples with a bin size of $\Delta \omega_{\rm lin} = 10\;\mathrm{Mpc}$. Finally, we compute the statistical metrics we introduced in \ref{subsec: stat_methods}. The result of this procedure applied to mocks is shown in the upper panel of Figure \ref{fig: test_mocks}. On the left, we show the median of the marginalised posterior on $A_{\rm lin}$, whereas, on the right, we show its 95\% credible interval. The light blue lines are the result for each mock, and the dark blue is the mean of 100 mocks. Notice that the credible interval on $A_{\rm lin}$ is approximately constant for $\omega_{\rm lin} \approx [100, 500]\;\mathrm{Mpc}$, but tends to increase for higher frequencies. What is causing this degradation is the window function convolution, as discussed in Section \ref{sec: window}. Also, note that in both left and right figures, there is an increase in noise at $\omega_{\rm lin} = 150\;\mathrm{Mpc}$, which arises when the primordial feature becomes approximately degenerate with the BAO oscillation. Another point worth mentioning is that the 95\% credible interval seems to saturate for $\omega_{\rm lin} \approx 4000\;\mathrm{Mpc}$. It happens because the window function degradation cannot occur perpetually, since we are also imposing a uniform prior on the amplitude (see Table \ref{tab: PF_all_priors}). At some point, this degradation saturates the prior, and, consequently, the scatter of $A_{\rm lin}$ around the origin. We can note that the scatter in median $A$ values rises with frequency in a similar way to the increase in 95\% confidence limits, but that the occasional mode has a preferred value above that limit: this is exactly the look-elsewhere effect in operation, although the effect seems to be modest (see \S \ref{subsec: look-elsewhere} for a detailed discussion).

We now move to the results for logarithmic features. In this case, we scan for features with $\omega_{\rm log} \in [10, 360]$ and bin the resulting posterior using a bin size of $\Delta \omega_{\rm log} = 1$. The results are presented in the middle panel of Figure \ref{fig: test_mocks}. Each light orange line represents the outcome of a single mock, whereas the dark orange line represents the mean of 25 mocks. The window function degradation is present once again, but the way it acts on the amplitude differs from the case of linear oscillations (see the discussion at the end of Section \ref{sec: window}). For the frequency range considered here, the error is not expected to saturate the prior. The degeneracy between logarithmic features and the BAO is not as prominent as in the case of the linear feature.

Next, we consider step features. The results for this model are in the second-last panel of Figure \ref{fig: test_mocks}. We scan the range $\omega_{\rm step} \in [100, 800]\; \mathrm{Mpc}$, and bin the posterior with a bin size $\Delta \omega_{\rm step} = 10 \; \mathrm{Mpc}$. Although this model is made of linear oscillations, its error increases more rapidly than for the linear features. The reason for this is that according to Eqs.\,\ref{eqn: step_w0} and \ref{eqn: step_w1}, the overall oscillation amplitude is decreased as we move to higher values of $\omega_{\rm step}$. This, when added to the natural window function effect, limits the constraining power of this model. The Planck 2018 inflation paper quotes a best-fit step model with a frequency $\omega_{\rm step} \sim 1000$ \citep{Planck2018_inflation}. As we will see in the next section, the current LSS constraining power in this range of parameter space is quite poor.

Finally, for sound features, we scanned the range $|\tau_f| \in [75, 900]$ and binned the posterior with a bin size of $\Delta \tau_{f} = 10 \; \mathrm{Mpc}$. The prior on the other parameters are in Table \ref{tab: PF_all_priors}. The results are presented in the bottom panel of Figure \ref{fig: test_mocks}. As for the step feature, discussed in the previous paragraph, the degradation of this PF amplitude due to non-linear damping, window-function convolution, and envelope function poses a strong obstacle when searching for this feature in LSS datasets. Our forecast with the mocks indicates that, even at small frequencies, the $95\%$ credible interval on $\mathcal{B}$ almost recovers the prior for all values of $\tau_f$. We thus already expect the data to be non-informative, and the model comparison should be inconclusive, with $\ln B_{01} \approx 0$. This will be confirmed later on with BOSS and eBOSS data in Section \ref{sec:results}.

Before moving to constraints from the real data, we decided to carry on a last sanity check. Since the primordial feature analysis is an extension of an isotropic BAO pipeline, we studied whether we could recover BAO constraints present in the literature. Our measurements are presented in Table \ref{tab: BAO-only}. Starting with the BOSS high-$z$, we find the same median for $\alpha_{\rm iso}$ as quoted in the official analysis \cite{BOSS:2016hvq}, but with an error approximately 2\% larger. We also found a good agreement for BOSS low-$z$, with our error bar approximately $5\%$ larger. For eBOSS, we recover the same error as in the official analysis \cite{Neveux:2020voa} and the median of our result is 0.1$\sigma$ away from the official result. We, therefore, conclude our BAO pipeline is consistent with previous results.

\section{Large-Scale Structure constraints}
\label{sec:results}
In this section, we present the large-scale structure constraints on primordial features obtained by combining both BOSS and eBOSS data. For each model we compute the general metrics we introduced in Section \ref{sec: methods}. We scrutinize our results in Section \ref{sec: discussion}.
\begin{figure}[t!]
    \centering
      \includegraphics[width = \textwidth]{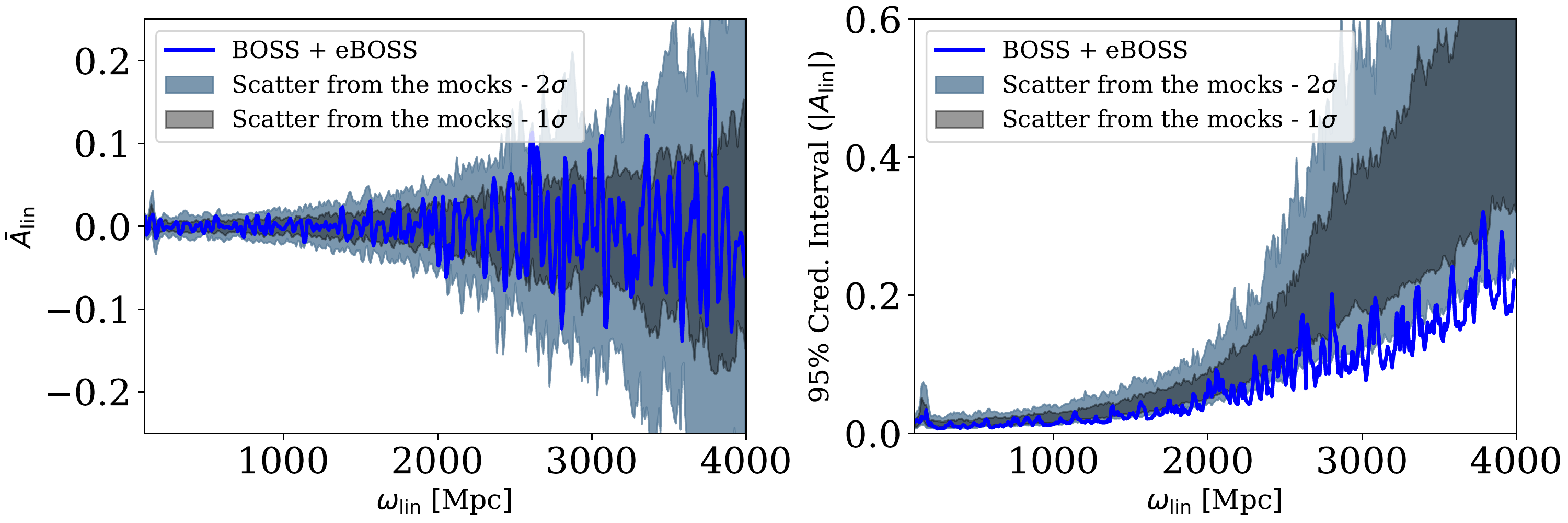}
    \includegraphics[width = 0.7\textwidth]{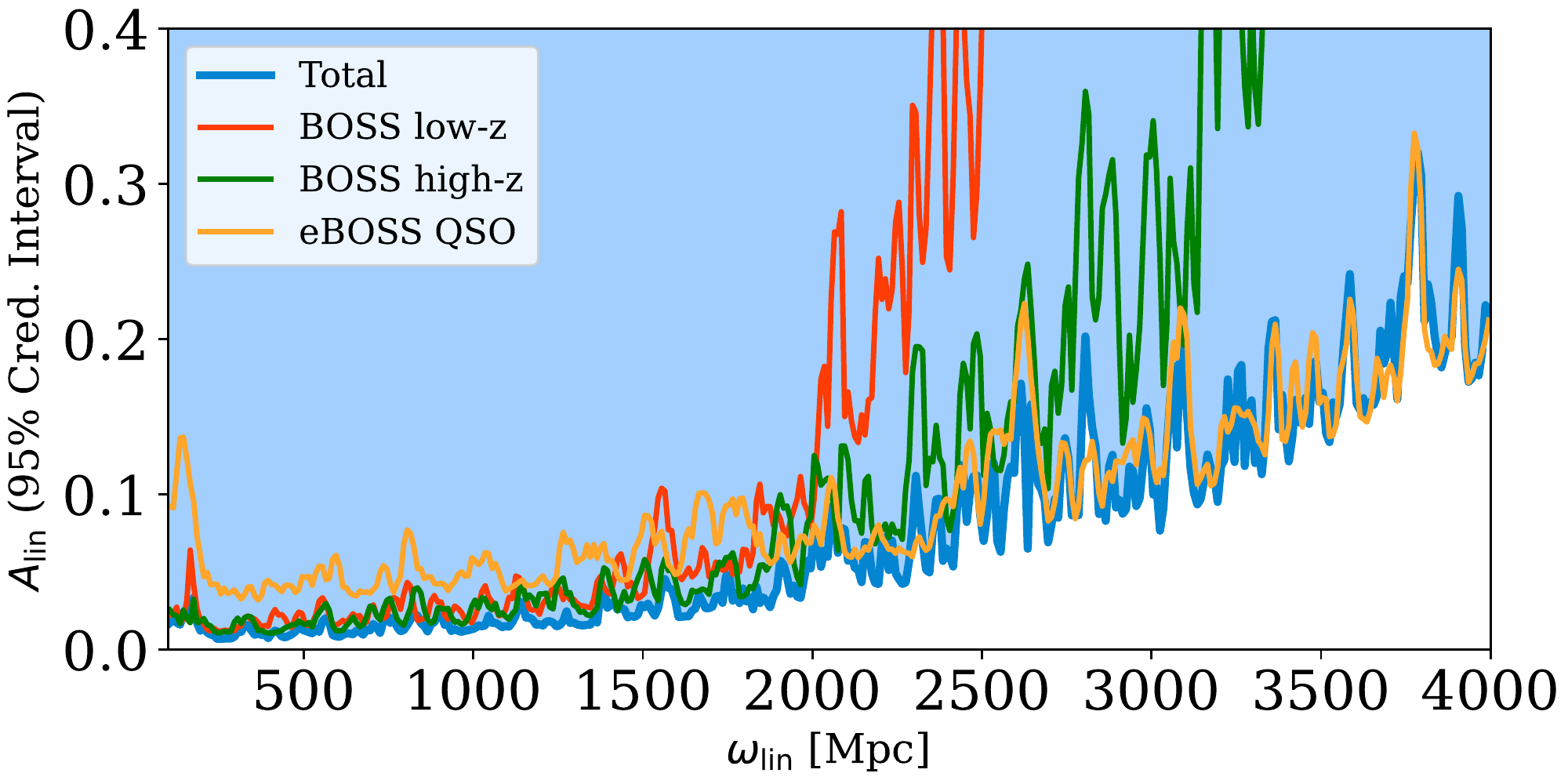}
    \caption{\textit{top-left}: Median of the marginalised posterior on $A_{\rm lin}$ for mocks and data. We also show the 1 and 2$\sigma$ regions from the scatter of the 100 mocks (light blue and grey backgrounds). \textit{top-right:} The 95\% credible interval on $A_{\rm lin}$ for both data and mocks. The background represent the same as before. \textit{bottom:} How each data product contributes to the combined result for the credible interval. The current LSS data discard the blue coloured region for linear features at 95\% confidence level.}
    \label{fig: lin_feature_result}
\end{figure}

\subsection{Linear features}

In Figure \ref{fig: lin_feature_result} we show the results for linear features. The upper-left panel shows the median of the marginalised posterior for $A_{\rm lin}$ as a function of the feature oscillation. Notice that we included BOSS constraints above the Nyquist frequency given in Table \ref{tab: products}, which is $\omega_{\rm lin} = 2323 \, \mathrm{Mpc}$. These were obtained by switching the binning of the BOSS power spectrum to be the same as for eBOSS. These results extend the range of the combined constraint, and they emphasize the different scaling of the error with the frequency. We use a bin size of $\Delta \omega_{\rm lin} = 10\,\mathrm{Mpc}$ starting with the upper left plot, where we see that although the observed medians for $A_{\rm lin}$ lie inside the scatter from the mocks, they are not as noisy, especially as we move to higher frequencies. A consequence is that the credible interval for the data should be tighter than almost all mocks, which is corroborated in the upper-right plot, where we compare the 95\% credible intervals. We conclude that the observed dataset is an outlier of the mocks, and such a discrepancy was reported previously in the literature \cite{beutler2019primordial}. We discuss the consequences of this discrepancy between data and mocks in the next section. In the lower panel, we show how each dataset contributes to the total constraint on $A_{\rm lin}$. At small frequencies, where the window function convolution is not predominant, the BOSS dataset dominates the constraint. At $\omega_{\rm lin} \approx 2500 \; \mathrm{Mpc}$ there is a cross-over between BOSS and eBOSS, and thereafter eBOSS dominates over BOSS. Finally, in the top panel of Figure \ref{fig: bayes_factors} we show the Bayes factor for linear PF as a function of their frequency.

\begin{figure}[t!]
    \centering
      \includegraphics[width = \textwidth]{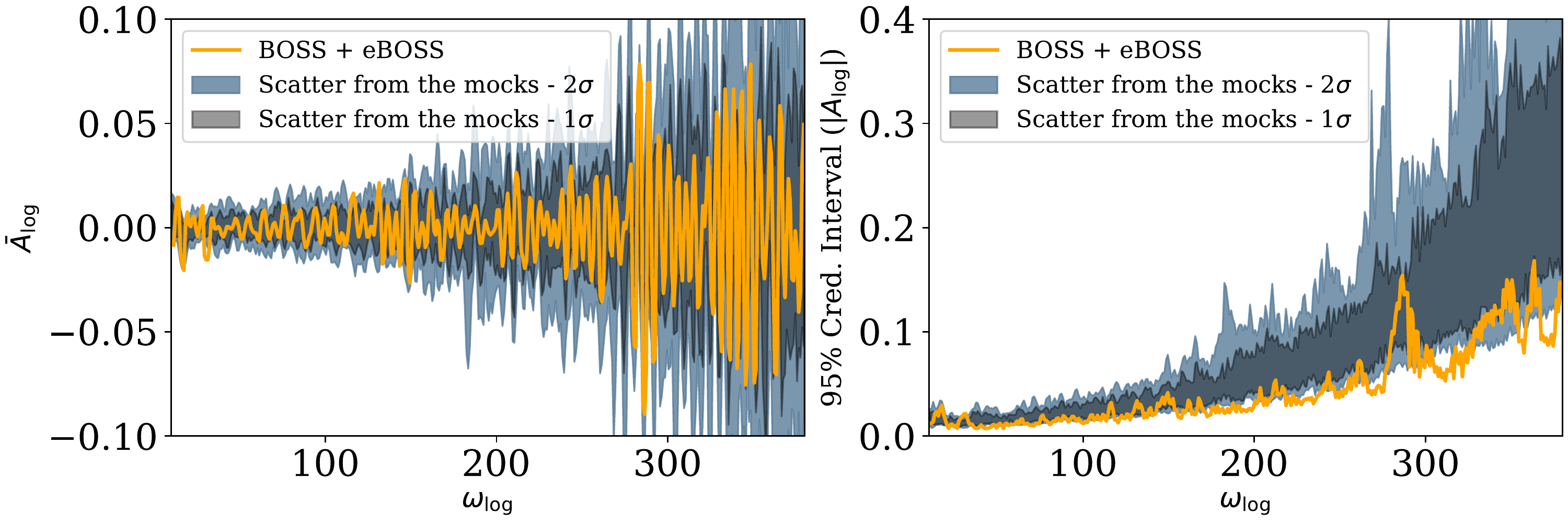}
    \includegraphics[width = 0.7\textwidth]{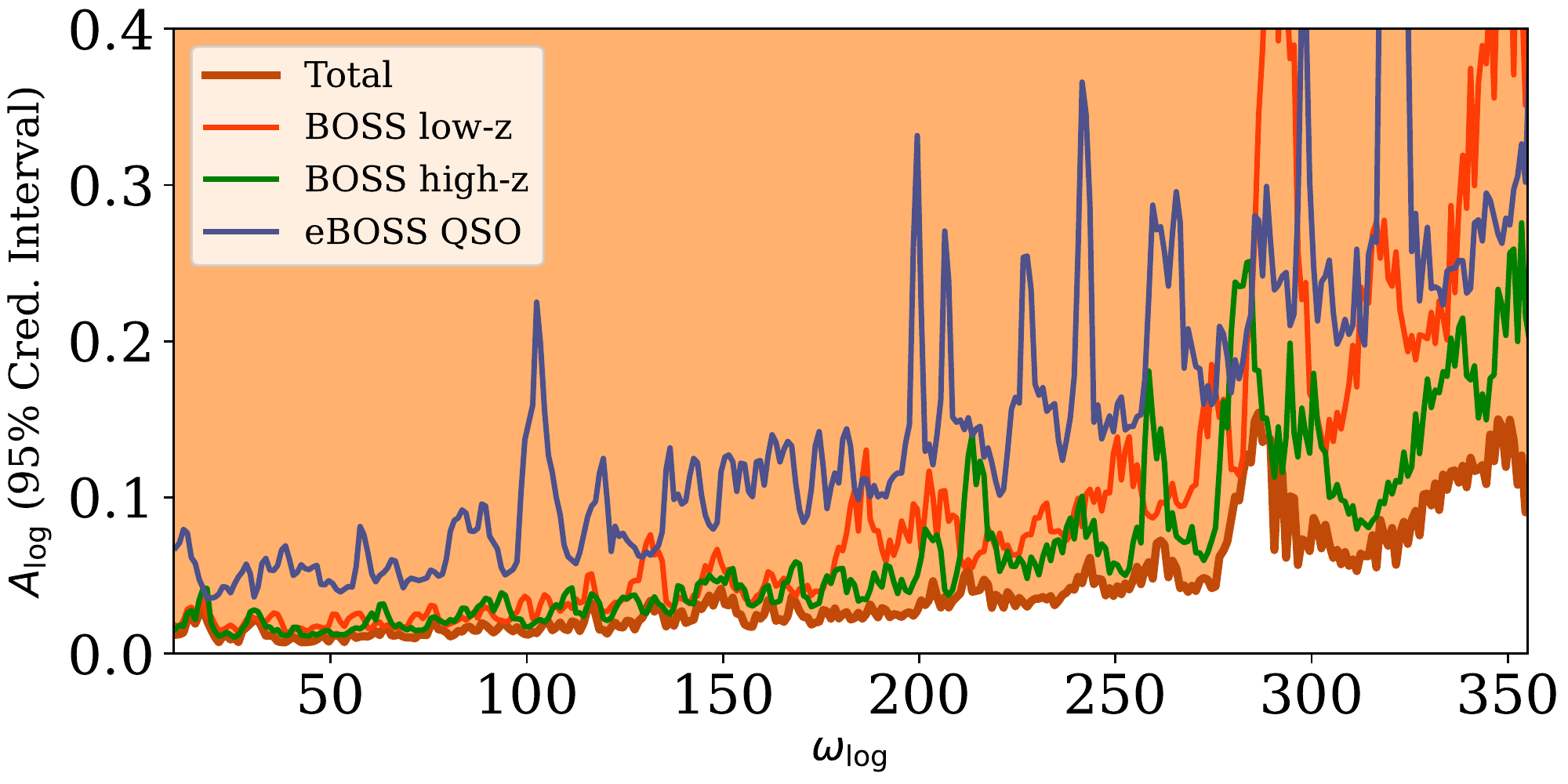}
    \caption{\textit{top-left}: Mean of the marginalised posterior on $A_{\rm log}$ for mocks and data. We also show the mean result of the mocks. We also show the 1 and 2$\sigma$ regions from the scatter of the 100 mocks (light blue and grey backgrounds) \textit{top-right:} The 95\% credible interval on $A_{\rm log}$ for both data and mocks. The background represent the same as before. \textit{bottom:} How each data product contributes to the combined result for the credible interval. The current LSS data discard the orange coloured region for logarithmic features at 95\% confidence level.}
    \label{fig: log_feature_result}
\end{figure}

\begin{figure}[t!]
    \centering
      \includegraphics[width = \textwidth]{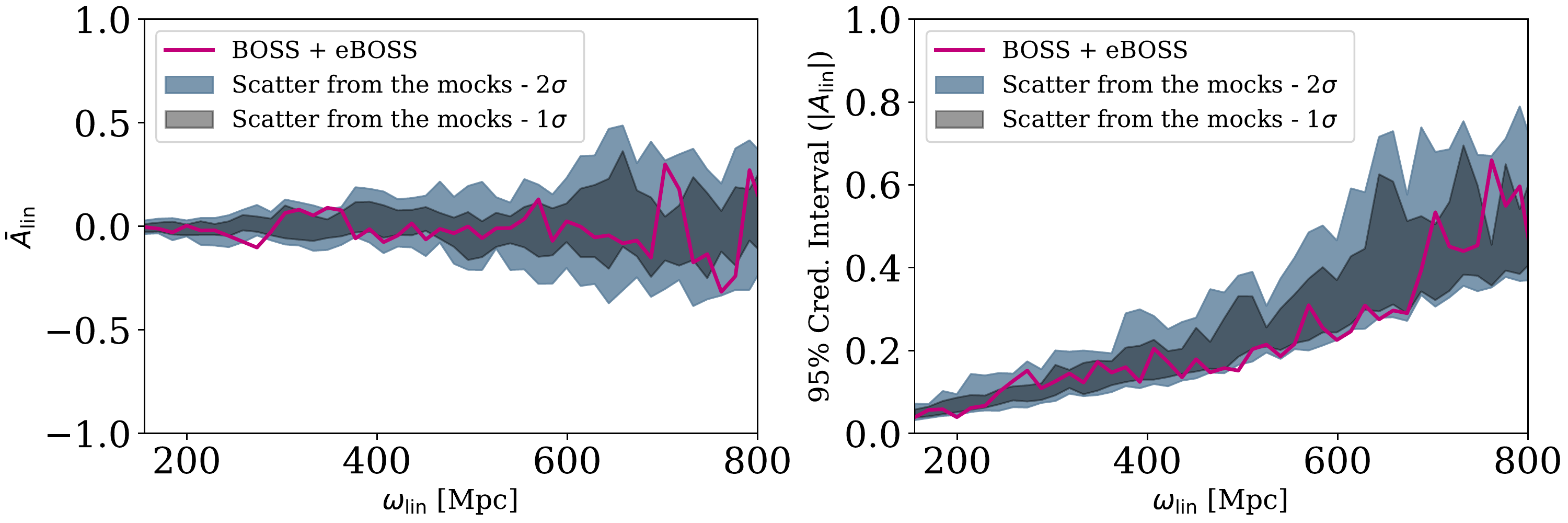}
    \includegraphics[width = 0.7\textwidth]{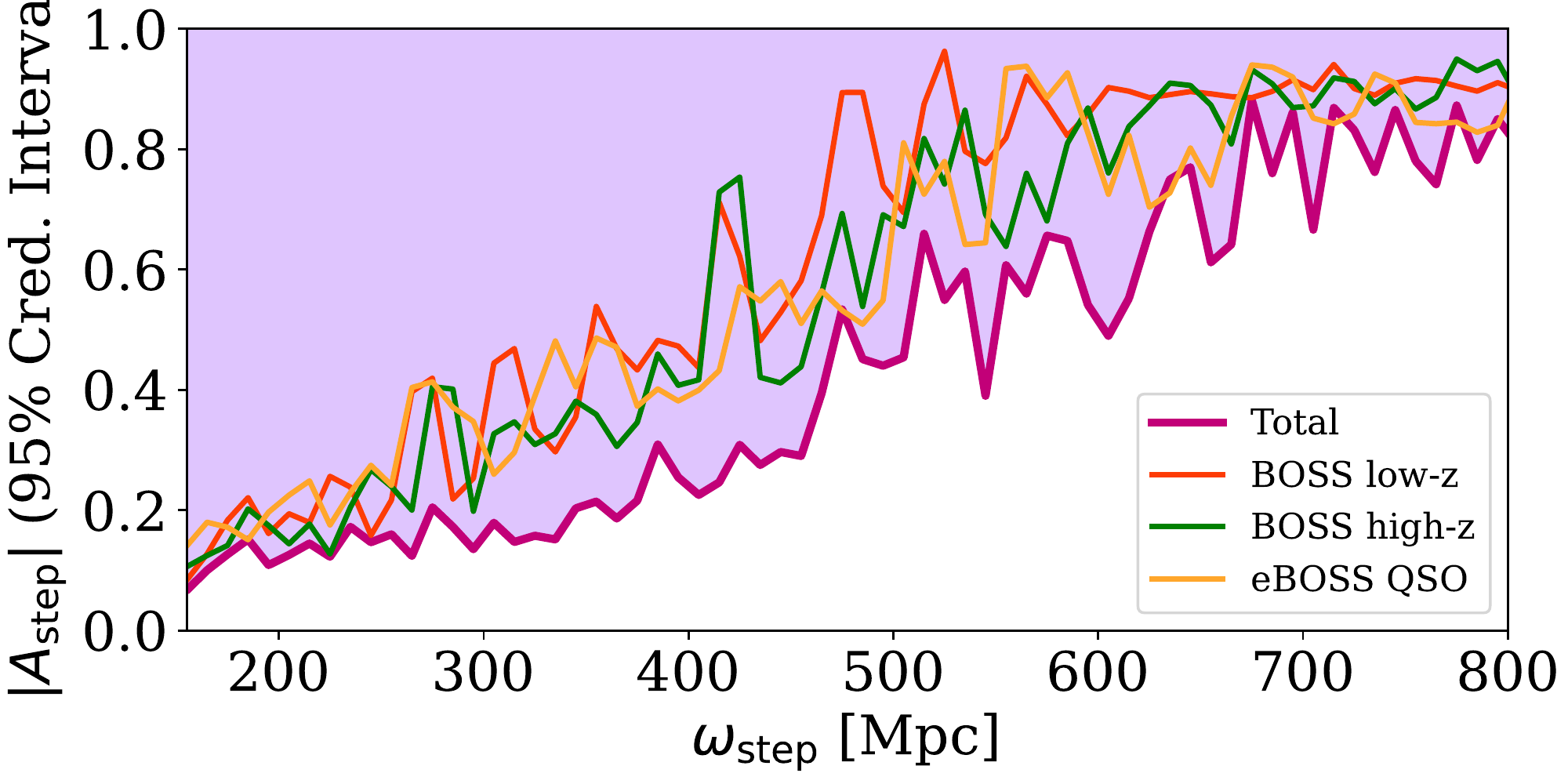}
    \caption{\textit{top-left}: Mean of the marginalised posterior on $A_{\rm step}$ for mocks and data. We also show the mean of the mocks as well as the 1 and 2$\sigma$ regions from the scatter of the 100 mocks (light blue and grey backgrounds) \textit{top-right:} The 95\% credible interval on $A_{\rm step}$ for both data and mocks. The background represents the same as before. \textit{bottom:} How each data product contributes to the combined result for the credible interval. The current LSS data discard the lavender coloured region for step features at 95\% confidence level.}
    \label{fig: step_feature_credible}
\end{figure}
\subsection{Logarithmic features}

In Figure \ref{fig: log_feature_result}, we show the results for logarithmic features. In the upper-left and upper-right panels, we show the median and the 95\% credible interval for $A_{\rm log}$, respectively, as a function of frequency. Here we used $\Delta \omega_{\rm log} = 1$. In the bottom panel, we show how each dataset contributes to the combined credible interval. There is no clear cross-over between BOSS and eBOSS, and both datasets are equally informative for $\omega_{\rm log} \sim 300$. There is a bump in the credible interval for $\omega_{\rm log} \sim 280$, which indicates that around this frequency the null hypothesis may be excluded with high significance. In Figure \ref{fig:chi2_log_feature}, we provide additional information about this peak. On the left panel, we show the $\Delta \chi^2$ relative to the vanilla inflation model for all datasets, and in the right panel, we show the marginalized posterior for that frequency bin. The Bayes factor for logarithmic features is shown in the middle panel of Figure \ref{fig: bayes_factors}.

\subsection{Step and sound features}

We show the large-scale structure constraints on step features in Figure \ref{fig: step_feature_credible}, following the same structure as linear and logarithmic features. For this result, we used a bin size of $\Delta \omega_{\rm step} = 10\,\mathrm{Mpc}$. For this PF model, data and mocks seem to agree well. All datasets have similar constraining power for this model, saturating the prior at the same frequency $\omega_* \approx 1230 \, \mathrm{Mpc}$. The Bayes factor for step features as a function of the frequency is shown in the bottom panel of Figure \ref{fig: bayes_factors}.

\begin{figure}
    \centering
    \includegraphics[width = 0.98\textwidth]{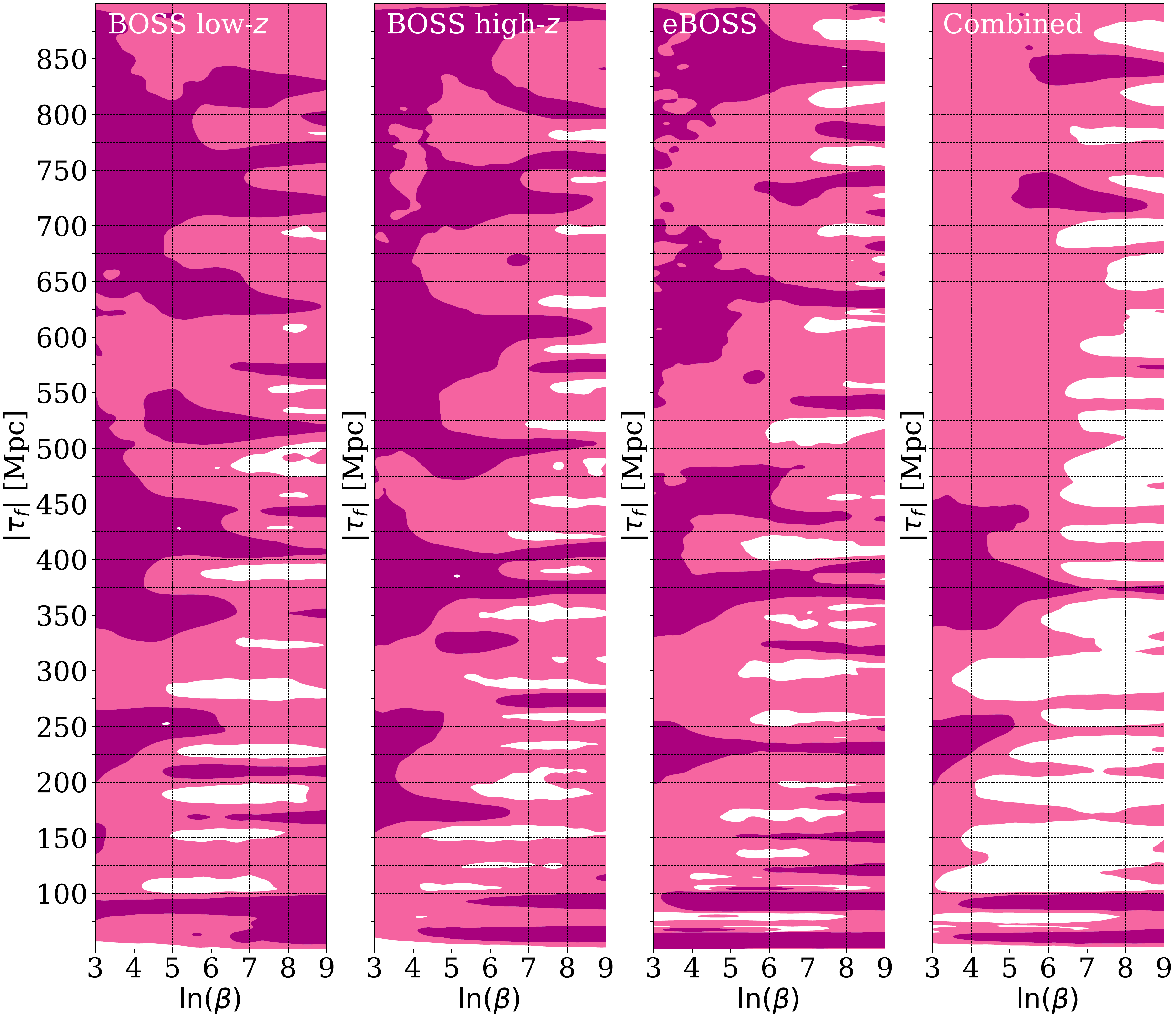}
    \caption{The $1\sigma$ and $2\sigma$ contours of the marginalized posterior on $\tau_f$ and $\ln(\beta)$ for sound features using the most recent LSS datasets. Although we have a weak constraint on the PF amplitude, $\mathcal{B}$, this plot shows preferred oscillatory modes (or preferred PF scale, $\tau_f$), which are the darker horizontal contours. The Bayes factor for this model obtained from BOSS low-$z$, high-$z$ and eBOSS Quasars are $\ln B_{01} = 0.620, 0.093, 0.045$, respectively. It is therefore neither favoured nor discarded by current LSS data.}
    \label{fig: sound_feature_modes}
\end{figure}

In Figure \ref{fig: sound_feature_modes}, we display the two-dimensional marginalized posterior for $\tau_f$ and $\ln \beta$ for each dataset and their combined result. Notice that the current LSS datasets have favoured regions in parameter space around some values of $\tau_f$, which are the darker horizontal contours in this figure.

\subsection{The look-elsewhere effect}
\label{subsec: look-elsewhere}
\begin{figure}[t!]
    \centering
    \includegraphics[width = 0.48 \textwidth]{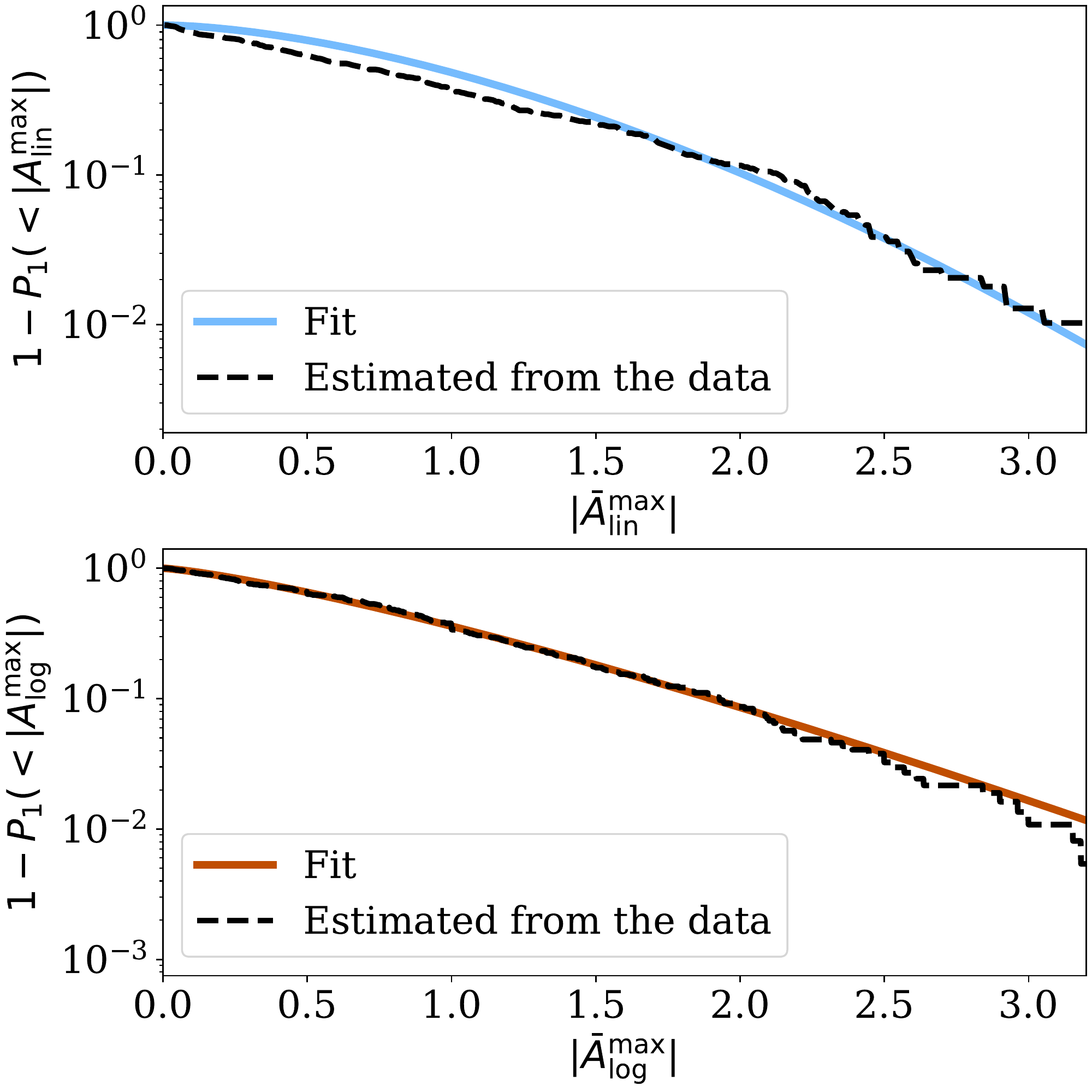}
    \includegraphics[width = 0.48\textwidth]{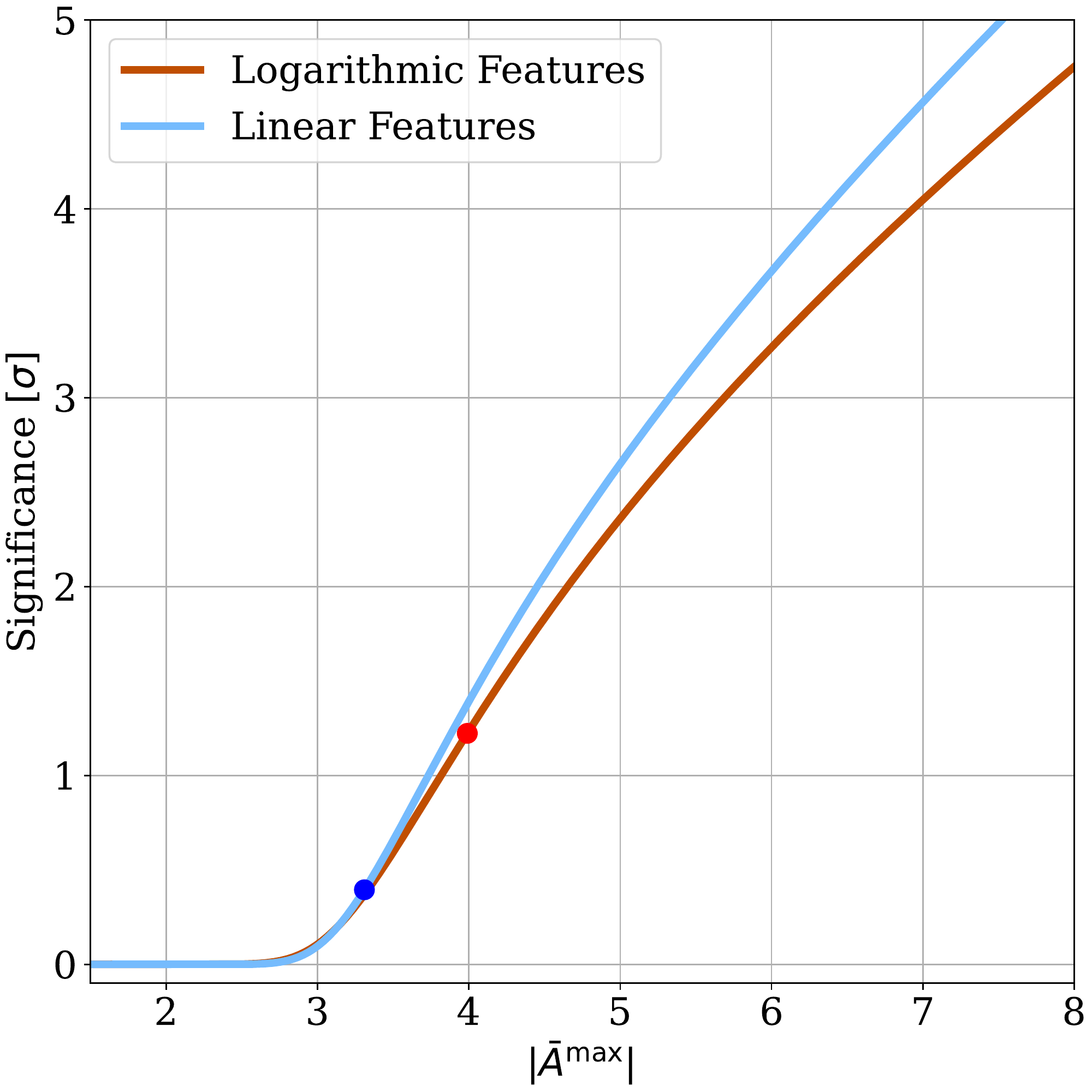}
    \caption{\textit{left}: The CDF of the maximum values of $|\bar{A}|$ for individual frequency channels for linear (upper) and logarithmic (bottom) feature. The black-dashed line is the CDF measured from the data, whereas the coloured lines are the best-fits of the template $\tilde{P}(<x) = 1 - \exp
\left[{-(x/b)^a}\right]$. The values of $(a,\;b)$ for linear and logarithmic features are $(1.61,\; 1.18)$ and $(1.44
,\;1.06)$, respectively; \textit{right}: How statistically relevant the events in the data are, at individual frequency bins, in light of the look-elsewhere effect. The red point indicate the $3.99\sigma$ event we see in the data for logarithmic features at $\omega_{\rm log} \approx 286$. When the look-elsewhere effect is taken into account, the significance of this event is reduced to to $1.22\sigma$. Similarly, the blue point indicate a $3.31\sigma$ event for linear features at $\omega_{\rm lin} \approx 2605\,\mathrm{Mpc}$. The look-elsewhere effect reduces its significance to $0.39\sigma$.}
    \label{fig:look_elsewhere}
    \end{figure}
In all of this, it is important to note that considering a number of different frequency bins makes our analysis susceptible to the `look-elsewhere' effect \cite{Gross:2010qma,Bayer:2020pva}. In other words, an analysis of real data might indicate a preferred value of $|A|$ for some frequency that is well above the 95\% limit derived from analyzing mocks -- but this need not constitute a true detection of primordial features if there was no reason to focus on that frequency in advance. The more frequency channels we inspect, the more likely it is to have one with an outlying value of $|A|$. In light of this effect, the statistical significance of events at a specific frequency is reduced. The impact of the look-elsewhere effect can be understood by analyzing the distribution of the maximum amplitudes that can be measured under the null hypothesis, 
\begin{equation}
|\bar{A}|^{\max }=\max _{\phi, \omega} |\bar{A}|\left(\omega, \phi\right)\;,
\end{equation}
where $\bar{X} \equiv \hat{X}/\rm{Var}(X)^{\frac{1}{2}}$. We are thus interested in computing 
\begin{equation}
P_{N_{\rm eff}}\left(|\bar{A}|^{\max } \geq x\right)=1-P_{N_{\rm eff}}\left(|\bar{A}|^{\max }<x\right) = 1-\left[P_1\left(|\bar{A}|^{\rm max}<x\right)\right]^{N_{\rm eff}}\;,
\label{eqn: look-elsehwere}
\end{equation} 
where $N_{\rm eff}$ is the effective number of independent frequency channels, $P_{N}\left(|\bar{A}|^{\max }<x\right)$ is the cumulative distribution function (CDF) of the distribution of maximum $|\bar{A}|$ over a total of $N$ frequency channels, and $P_1\left(|\bar{A}|^{\max }<x\right)$ is the same but for a single frequency bin. The safest way to evaluate Eq.\,(\ref{eqn: look-elsehwere}) is to have an ensemble of reliable mocks with no PF, from which we can construct the probability distribution function (PDF) of $|\bar{A}|^{\rm max}$, and consequently compute its CDF, $P(|\bar{A}|^{\rm max}<x)$, to be used in Eq.\,(\ref{eqn: look-elsehwere}). But unless the mocks are completely realistic, this approach may not be reliable; we therefore make a more indirect argument, focusing on the effective number of independent frequency bins, $N_{\rm eff}$.

To estimate $N_{\rm eff}$, we measure both $P_N$ and $P_1$ from an ensemble of mocks, plug them into the last equality of Eq.\,(\ref{eqn: look-elsehwere}) and solve for $N_{\rm eff}$. For linear and logarithmic features, we found that $N_{\rm eff} \approx 220$ and $215$, respectively. The associated effectively independent frequency bins are $\Delta \tilde{\omega}_{\rm lin} \approx 17.70\;\mathrm{Mpc}$ and $\Delta \tilde{\omega}_{\rm log} \approx 1.72$, both a little larger than the adopted values of $\Delta \omega_{\rm lin} = 10\;\mathrm{Mpc}$ ($N=390)$ and $\Delta \omega_{\rm log} =1$ ($N=370$), respectively. 

\begin{figure}[t!]
    \centering
    \includegraphics[width = 0.95\textwidth]{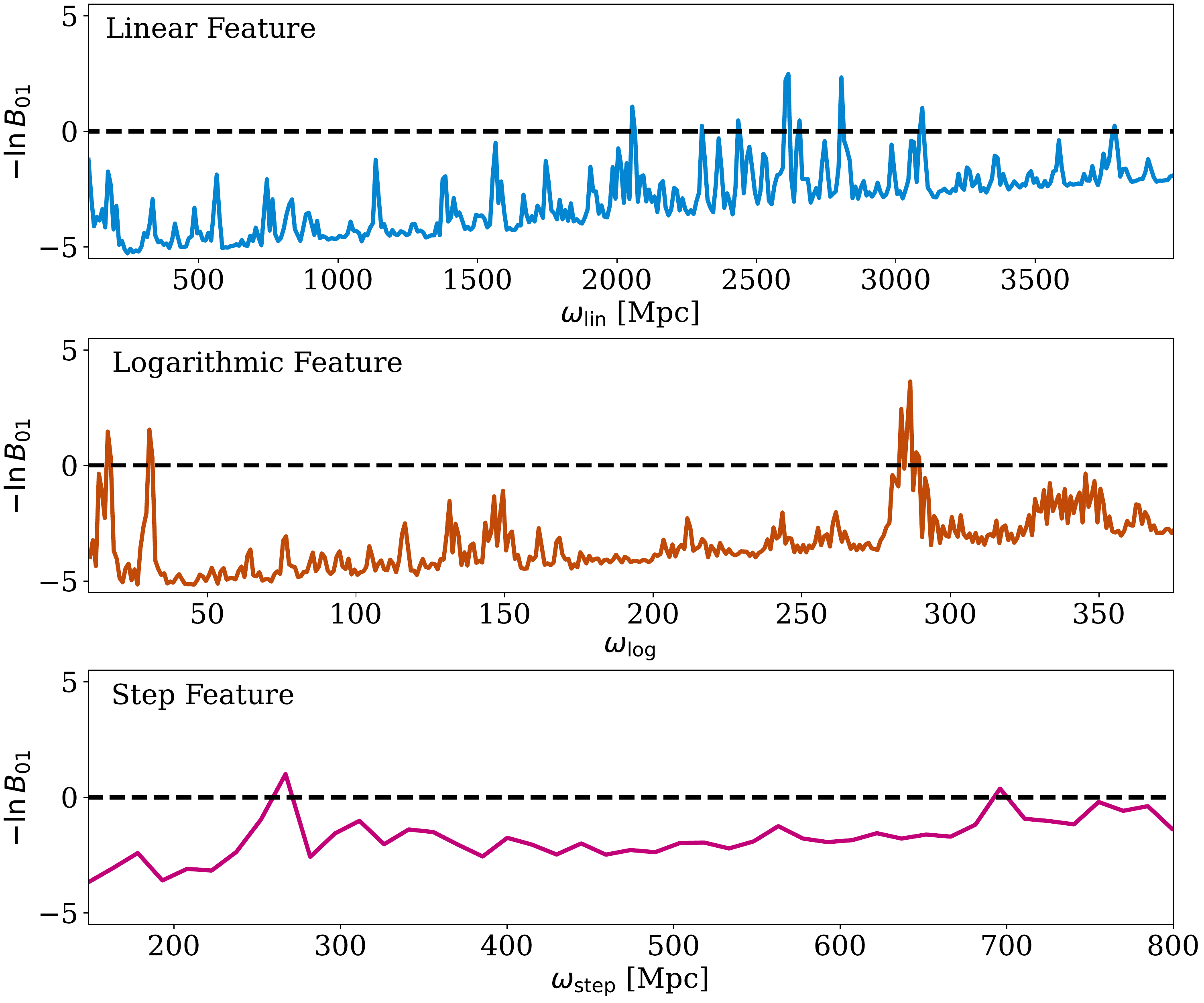}
    \caption{The combined Bayes factor (from BOSS and eBOSS datasets) for linear, logarithmic, and step features. Peaks above the black dashed line favour the PF model over the vanilla inflationary model in that frequency band. The opposite holds for points below the line. According to Jeffreys' scale \cite{jeffreys1998theory}, moderate and strong evidence is associated with $|\ln B_{01}| \approx 2.5$ and $5.0$, respectively. Apart from a single frequency bin for logarithmic features, the current LSS data strongly favour the vanilla inflationary scenario over the PF signals we studied in this work. We used the following bins for the frequencies: $\Delta \omega_{\rm lin} = 10\,\mathrm{Mpc}$, $\Delta \omega_{\rm step} = 10\,\mathrm{Mpc}$ and $\Delta \omega_{\rm log} = 1$. For sound features the posterior is dominated by the prior, $\ln\left(B_{01}\right) \approx 0$ for all $\tau_f$, hence we decided to omit this model from this study of the Bayes factor.}
    \label{fig: bayes_factors}
\end{figure}

Now, as shown in Appendix \ref{app: non-linear impact}, the eBOSS and BOSS mocks are not a perfect representation of the data, and they have a noisier constraint on the amplitudes of PF and thus a different single-channel amplitude distribution, $P_1(|A|)$. We did however note that $P_1(|\bar{A}|<x)$ for both mocks and data is well-fitted by the form $\tilde{P}(<x) = 1 - \exp
\left[{-(x/b)^a}\right]$. 
Our approach is therefore to use the $P_1$ function measured from data but to allow for the look-elsewhere effect by using the $N_{\rm eff}$ values quoted above, obtained from the mocks. We believe this to be a defensible procedure, since the degree of correlation between frequency channels is set by the survey window, and should not depend on the detailed nature of the mock fluctuations. For convenience, we convert the $p$-value computed using Eq.\,(\ref{eqn: look-elsehwere}) into 
the corresponding $S-\sigma$ point of a Gaussian (via $S=2^{1/2} \operatorname{Erf}^{-1}[1-p]$,  where $\operatorname{Erf}^{-1}$ is the inverse error function).

In Figure \ref{fig:look_elsewhere}, in the left panel, we show how well $P_1(|A|<x)$ measured using the data is fitted by $\tilde{P}$. We use this to map the real significance of events in the data at individual frequency bins, and we present the result in the right panel of Figure \ref{fig:look_elsewhere}. The $x$-axis in that figure represents the values of $|A|$ normalized to its standard deviation, and in the $y$-axis we show its real significance when the look-elsewhere effect is taken into account. We also included two points representing the strongest event we detected in the data for linear and logarithmic features. Note how the look-elsewhere effect reduces the significance of single-channel events.
\begin{table}[]
\centering
\begin{tabular}{|cccc|}
\hline
\multicolumn{1}{|l}{}               & \multicolumn{1}{l}{} & \multicolumn{1}{l}{$\Sigma_{\rm nl}\,[h^{-1}\,\mathrm{Mpc}]$} & \multicolumn{1}{l|}{} \\ \hline
\multicolumn{1}{|c|}{}              & BOSS low-$z$         & BOSS high-$z$                         & eBOSS Quasars         \\ \hline
\multicolumn{1}{|c|}{Mean of mocks}          &  $6.00 \pm 0.26$                  & $5.66 \pm 0.27$                                     & $13.57 \pm 1.46$                   \\
\multicolumn{1}{|c|}{Data} &$4.1^{+1.6}_{-1.3}$                     & $2.6^{+1.3}_{-1.8}$                                     & $2.44^{+0.94}_{-2.2}$                    \\ \hline
\end{tabular}
\caption{The value of $\Sigma_{\rm nl}$ for each dataset compared to their respective mocks. For each dataset, we performed a total of 100 fits from the mocks and a single one from the data. The figure we quote above for the mocks is the average best fit and its standard deviation, and for the data, we quote the $1\sigma$ range obtained from the posterior. For all datasets the value fitted from the mocks is larger than the data. This inconsistency between the mocks and the data poses problems with the interpretation of our results (see \S\ref{sec: discussion} for more details).}
\label{tab: non-linear damping}
\end{table}
\section{Discussion}

\label{sec: discussion}
In this section, we comment on the results presented in Section \ref{sec:results}. First, we scrutinize the comparisons between data and mocks and explain why they do not have similar constraints for global oscillatory features. Finally, we comment on the credible intervals and the Bayes Factors we obtained for each model, which constitutes our main results.

The results for linear and logarithmic features in Figures \ref{fig: lin_feature_result} and \ref{fig: log_feature_result}, respectively, show that the observed dataset is an outlier of the mocks. In \cite{beutler2019primordial}, the authors found the same trend, and the reason is the same as in the present work: mock catalogues prefer higher values for the non-linear damping parameter, $\Sigma_{\rm nl}$. This leads to a higher scatter of the mean amplitude and a more substantial error. In Table \ref{tab: non-linear damping}, we compare the best-fit values of $\Sigma_{\rm nl}$ between data and mocks. The error we quote for mocks is the sample standard deviation of the best fit of the 100 mocks we analyzed. We see that the observed result is $\sim5\sigma$ away from the mocks for BOSS, and we find similar results for eBOSS. It reveals that the non-linear damping in the mocks is not consistent with the data. These mocks were generated using approximate methods, so we speculate that it may have induced non-physical smearing of features, probably due to some resolution issue. For BOSS Patchy mocks, the problem is less severe (see also the discussion in \S8.3 of \cite{BOSS:2016hvq}). Therefore, it explains why the disagreement we see in Figure \ref{fig: lin_feature_result} is not that extreme for low PF frequencies, which is the region in parameter space dominated by BOSS. In Appendix \ref{app: non-linear impact}, we show that if $\Sigma_{\rm nl}$ is fixed to the best-fit extracted from the data or N-body simulations, the PF amplitudes and errors fitted from the mocks are consistent with the data for all frequencies.

Another trend in our results is the increase of the error on the PF amplitude with its frequency. As we discussed in \S\ref{sec: window}, this is caused by the window function convolution, which will smear out any oscillatory signal present in the data. The smearing will be stronger for higher frequencies. Moreover, since the cosmic volume of the three datasets we used are different, the window function impact will be different. This effect is perceptible at the bottom panel of Figures \ref{fig: lin_feature_result}, \ref{fig: log_feature_result} and \ref{fig: step_feature_credible}, where we show how each dataset contributes to the combined constraint. Notice that this effect completely deteriorates the BOSS results for linear features at higher frequencies, causing the eBOSS QSO sample to dominate over BOSS.

In Figures \ref{fig: lin_feature_result}, \ref{fig: log_feature_result} and \ref{fig: step_feature_credible}, we also show the median and credible interval on the PF amplitude for the linear, logarithmic, and step models. These are the combined results of BOSS and eBOSS datasets. The coloured regions in the credible interval plots indicate the volume in parameter space excluded by current data at 95\% significance. In Figure \ref{fig: bayes_factors}, we show the Bayes factor, $-\ln B_{01}$, for these three models as a function of the frequency, which is the main result of this work. The Bayes factor is interpreted as the odds ratio between two models \footnote{For instance, $B_{01} = 2$ means that the model $\mathcal{M}_0$ is twice as likely to be true relative to $\mathcal{M}_1$, given the data.}. Peaks above the dashed line in Figure \ref{fig: bayes_factors} indicate that current data prefer the PF over the vanilla model, and the opposite holds for points below that line. The significance of these points can be interpreted according to the Jeffreys' scale \cite{jeffreys1998theory}: moderate and strong evidence is associated with $|\ln B_{01}| \approx 2.5$ and $5.0$, respectively. Note that $B_{01}$ asymptotically approaches unity for high frequencies, a trend that will be present for all PF models. It happens because, for all oscillatory PF models, there is an associated limiting frequency that can be probed with a given cosmic volume, $\omega_*$. Therefore, the error on the PF amplitude, $\sigma\left(|A|\right)$, will increase indefinitely as $\omega \rightarrow \omega_*$ . In this regime, the data become less informative and consequently the posterior starts to be dominated by the prior, making the model comparison inconclusive \cite{Trotta:2005ar}.

Starting with linear features, in the upper panel of Figure \ref{fig: bayes_factors}, we see that apart from some specific bins of $\omega_{\rm lin}$ the featureless inflationary model is favoured over the linear feature model. According to Jeffreys' scale, all the significant events support the simple inflationary model. The data have four 3$\sigma$ events and no 4$\sigma$ events. None of the 3$\sigma$ events are in the range $\omega_{\rm lin} \leq 900 \;\mathrm{Mpc}$, in agreement with \cite{beutler2019primordial}. Regarding the events at larger frequencies, we find that approximately 20\% and 7\% of the mocks have at least one 3$\sigma$ and 4$\sigma$ event, respectively. It indicates that the peaks we see are not statistically significant, since we see similar events in the mocks with modest frequency.
\begin{figure}
    \centering
    \includegraphics[width = 0.5\textwidth]{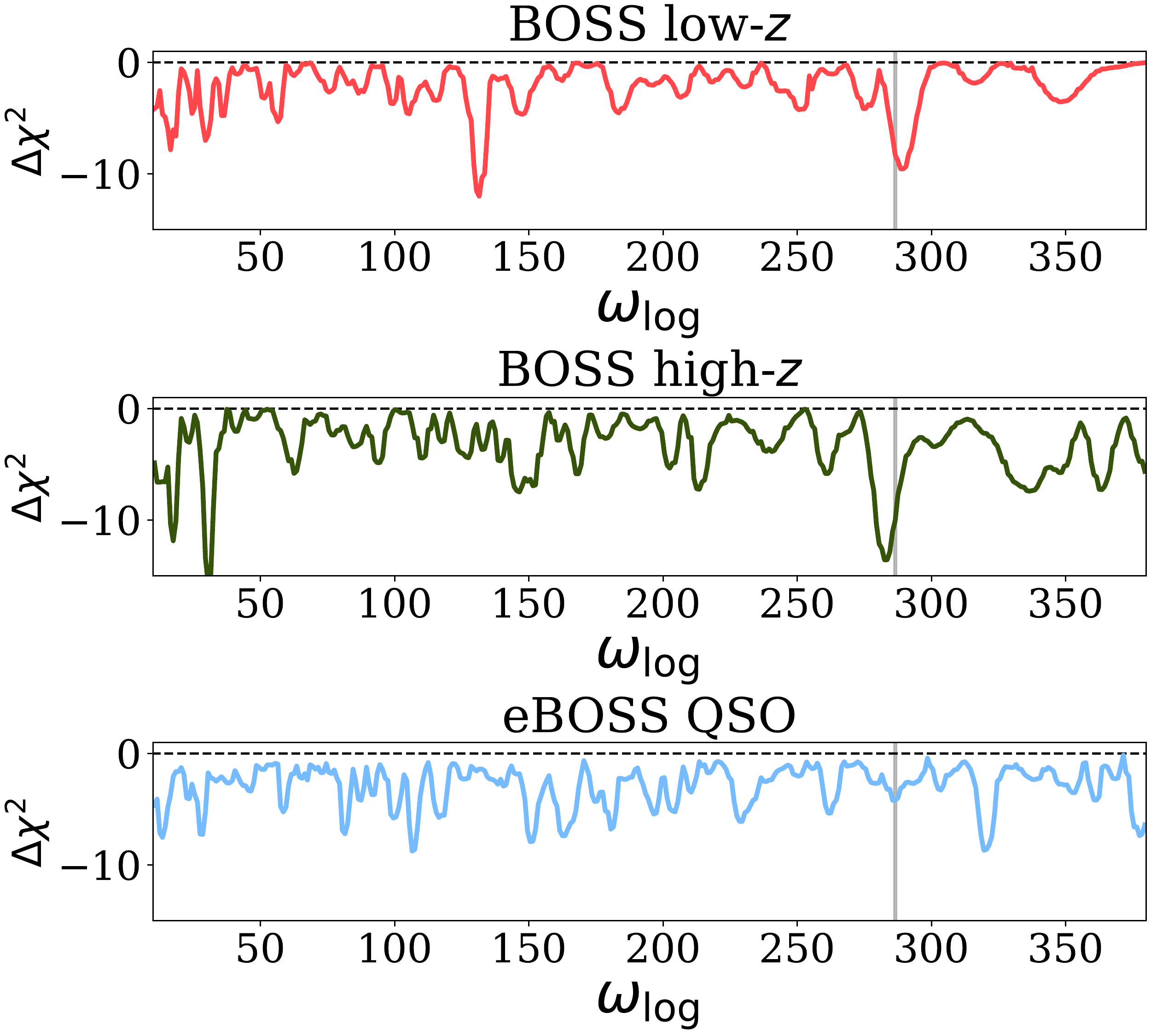}
    \includegraphics[width = 0.48 \textwidth]{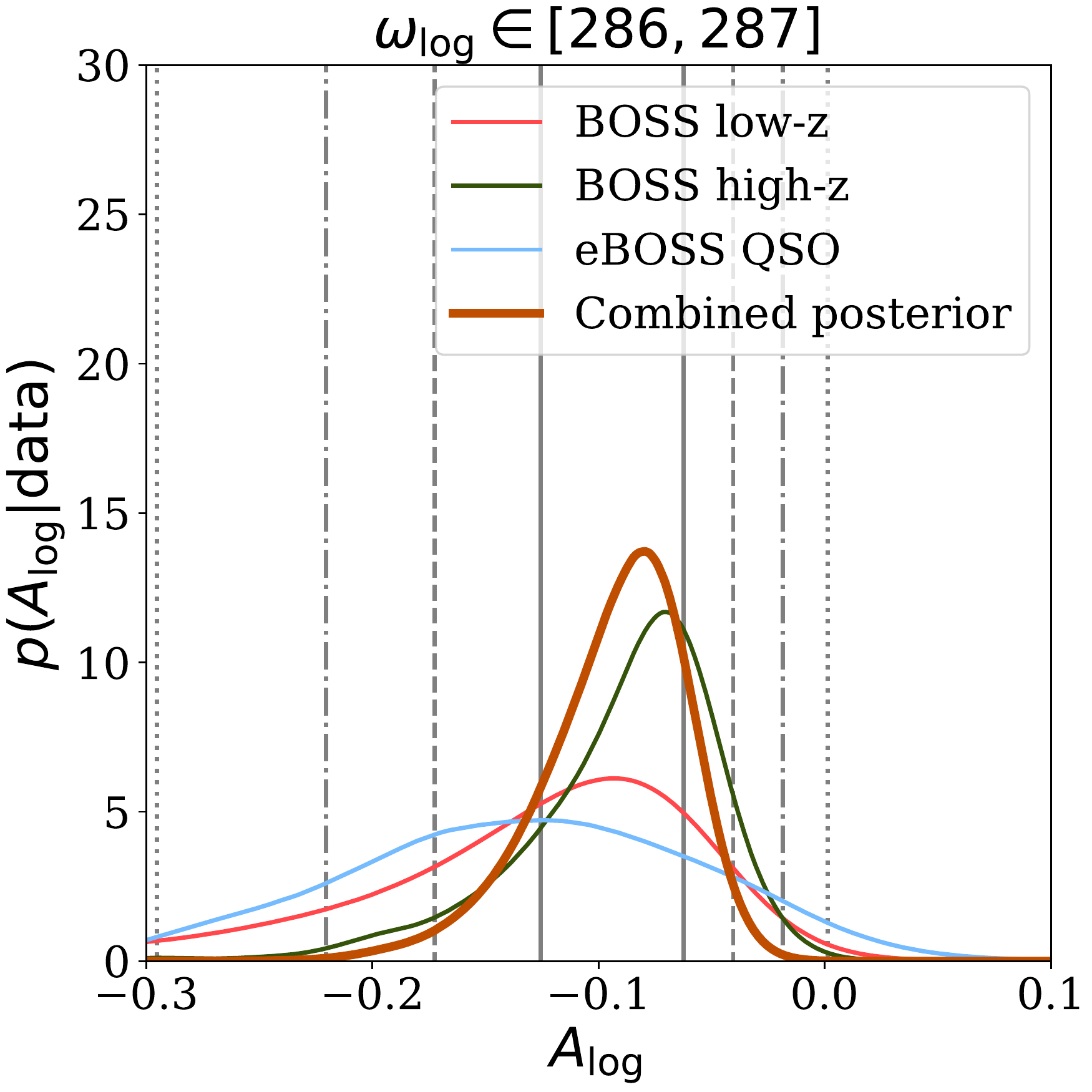}
    \caption{\textit{left panel:} The $\chi^2$ profile for logarithmic features in the range $\omega_{\rm log} \in [10,380]$. The black solid line is the value for the featureless model. The more striking downward peaks at $\omega_{\rm log} \approx 286$ for the BOSS data sets are associated with the combined (BOSS + eBOSS) $\ln B_{01} \approx -4$ feature in Figure \ref{fig: log_feature_result}. \textit{right panel:} The marginalised posterior for $A_{\rm log}$ in the frequency bin $286 < \omega_{\rm log} < 287$, where the logarithmic feature is favoured the most over the featureless model. The pairs of vertical lines represent the $1\sigma$, $2\sigma$, $3\sigma$, and $4\sigma$ limits, respectively.}
    \label{fig:chi2_log_feature}
\end{figure}
The Bayes factor for logarithmic features is shown in the middle panel of Figure \ref{fig: bayes_factors}. For $\omega_{\rm log} \sim 286$, we find $\ln B_{01} \approx -4$, indicating that for this frequency band, there is strong evidence of primordial logarithmic features. To investigate this event more deeply, we show in the left panel of Figure \ref{fig:chi2_log_feature} the $\Delta \chi^2$ profile between the featureless inflationary model and the logarithmic features models. The grey-shaded area represents the frequency range where the event occurs. The best-fit for $\omega_{\rm log} \in [286, 287]$ produces $\Delta \chi^2 = -8.32, -10.01, -4.20$ for BOSS low-$z$, high-$z$ and eBOSS, respectively. In the right panel of the same figure, we show the marginalized posterior for $A_{\rm log}$ with $\omega_{\rm log} \in [286, 287]$. The vertical lines there represent the $1\sigma$, $2\sigma$, $3\sigma$, and $4\sigma$ intervals around the peak of the posterior. The null hypothesis is discarded with a spurious significance of $\sim 3.9\sigma$. The real significance is obtained after the look-elsewhere effect is taken into account. Following the steps in \S\ref{subsec: look-elsewhere}, we conclude that this effect dilutes this significance to approximately $1.22\sigma$, which is far from allowing us to claim any detection, regardless of the substantial Bayes factor.

In the lower panel of Figure \ref{fig: bayes_factors} we show our results for step features. The vanilla model of inflation is favoured for all frequency bins. The Planck paper on inflation \cite{Planck2018_inflation} quotes a best-fit with $\omega_{\rm step} \approx 1230\;\mathrm{Mpc}$, but we cannot reach frequencies close to this value with current data. 

The last model we comment on is the sound feature. The Bayes factors for this model are $\ln B_{01} = 0.620, 0.886$, and $0.045$ for BOSS low-$z$, high-$z$, and eBOSS Quasars, respectively. According to Jeffreys' scale, these values indicate that the model comparison is inconclusive -- sound features are neither discarded nor preferred by current LSS datasets. The best fits for these same datasets lead to an improvement of $\Delta \chi^2 = -6.81, -8.76, -11.15$. Finally, in Figure \ref{fig: sound_feature_modes}, we show the marginalized posterior for the characteristic time scale of this PF, $\tau_f$, and its sharpness, $\ln\beta$ (see Eq. \ref{eqn: sound_spectrum}). A takeaway from this figure is that the current LSS data have a series of preferred modes in the posterior (darker horizontal contours). This is an improvement over previous results using the WiggleZ data \cite{Hu:2014hra}. Currently, the most stringent constraints for sound features are from the CMB \cite{Achucarro:2014msa, Achucarro:2013cva, Canas-Herrera:2020mme}, and since LSS alone can currently restrict regions in the $\tau_f$ parameter, regardless of not having strong constraints on the amplitude, a combined analysis would be ideal. We leave this task for future work.

\section{Summary and conclusions}
Understanding the nature of inflation is one of the major open questions in modern cosmology. To address this problem, many theories and frameworks were developed (e.g. see \cite{chluba2015features}) over recent decades. These extended models usually predict additional corrections to the power spectrum of curvature perturbations beyond the results of single-field slow-roll inflation models. Therefore, detecting any secondary signal lurking in the data would be a window into new physics happening during the early universe. In this paper, we obtained combined large-scale structure constraints on these additional signals, called primordial features (PF), using the BOSS DR12 and eBOSS QSO datasets. In \cite{beutler2019primordial} it was shown that LSS can lead to tighter constraints on PF than Planck, so we expect our results to be close to the optimal constraints that can be obtained with current cosmological datasets. They also showed that next-generation galaxy surveys such as DESI will improve the sensitivity to features by a factor of 7. Furthermore, the fact that the DESI quasar sample will have a larger cosmic volume will also extend the frequency range of linear features by a factor of 2.5.

We implemented a pipeline to search for PF signals considering an approximate method for the non-linearities present in the late-time cosmological probes, as discussed in Section \ref{sec: clustering_template}. This framework proved to be versatile and robust since it can be used to search for any oscillatory PF and is a simple extension of a typical BAO analysis. We used it to search for four different signals of PF, introduced in Section \ref{sec: primordial_features}: two phenomenological oscillatory templates (linearly and logarithmically spaced oscillations); PF due to the presence of a step in the inflaton's potential; and PF due to a variation in the inflaton's speed of sound. In Section \ref{sec: methods}, we presented the statistical metrics we used to compute the significance of any potential detection and applied these to mocks. This exercise revealed that the mocks we used are not a perfect reproduction of the data -- they prefer higher values of the non-linear damping, $\Sigma_{\rm nl}$ (see Table \ref{tab: non-linear damping} and Appendix \ref{app: non-linear impact}). 
We were able to circumvent this problem when estimating the impact of the look-elsewhere effect, as discussed in \S \ref{subsec: look-elsewhere}, which allowed us to estimate the global significance of the events we see in the data.

For linear features, we scanned the range $\omega_{\rm lin} \in [100, 4000] \; \mathrm{Mpc}$, and our results are presented in Figure \ref{fig: lin_feature_result}. We showed that although BOSS DR12 has a higher number density than the eBOSS QSO sample, the latter can lead to stronger constraints at high frequencies ($\omega_{\rm lin} > 2500 \; \mathrm{Mpc}$) due to the smaller fundamental mode in eBOSS.
For logarithmic features, we scanned the range $\omega_{\rm log} \in [10, 380]$, and the results are given in Figure \ref{fig: log_feature_result}. We found one event with a Bayes factor of $\ln B_{01} \approx -4$, which apparently favours the feature model relative to a featureless model. The null hypothesis for this frequency bin is eliminated with a significance of $3.9\sigma$. However, when the look-elsewhere effect is taken into account, the significance is reduced to approximately $1.22\sigma$. The results for step features are given in Figure \ref{fig: step_feature_credible}, and we found no significant indication that the data prefer this model with respect to the featureless inflationary model. In Figure \ref{fig: bayes_factors} we summarize the Bayes factor for all these three models as a function of their frequency. Finally, for sound features, we found global Bayes factors of $\ln B_{01} = 0.620, 0.886$ and $0.045$ for BOSS low-$z$, high-$z$, and eBOSS Quasars. According to the Jeffreys' scale, these models are thus neither favoured nor discarded by current LSS datasets alone. Nevertheless, in Figure \ref{fig: sound_feature_modes} we showed that LSS constraints have preferred values for the typical scale of the speed of sound reduction, $\tau_f$. The CMB constraints for this model also present a set of preferred values of $\tau_f$, so combining both LSS and CMB constraints will lead to a significant reduction in the allowed parameter space.

With this work, we are one step closer to extracting the maximum possible information about the primordial universe using  large-scale structure, leading to constraints competitive with (or, in some cases, surpassing) the CMB. Our approach provides a baseline, which we will seek to improve before applying it to the forthcoming DESI Y1 data. From the theoretical side, for instance, we plan to search for other primordial feature models and use more realistic modelling for the non-linear damping of features. Nevertheless, in all of this, the most thrilling improvement will be in the data itself, with a number density of objects being one order of magnitude larger, populating a volume approximately five times larger \cite{DESI:2022gle,Hahn:2022dnf,Raichoor:2022jab,Chaussidon:2022pqg}. This will ultimately lead to an improvement in the precision of PF amplitudes by a factor of $\sim\,$7 (see Figure 10 of \cite{beutler2019primordial}) and, perhaps more importantly, allow us to scan a frequency range $\sim\,$2.5 times larger, up to $\omega_{\rm lin} \sim 10\,000 \, \mathrm{Mpc}$.

\acknowledgments
This project has received funding from the European Research Council (ERC) under the European Union’s Horizon 2020 research and innovation program (grant agreement 853291). FB is a Royal Society University Research Fellow.
\appendix

\section{Comparing the parametrisation}
\label{app: phase_impact}
In the literature, there are results considering two distinct modelling approaches of the oscillatory features. In this appendix, we scrutinize their differences and how to correctly impose a non-informative prior. 

In \cite{Planck2018_inflation}, the search for linear and logarithmic features is performed assuming
\begin{equation}
    \delta P = A_{\rm X} \sin(\omega_{\rm X}k + \phi)\;,
\end{equation}
where $X = \mathrm{lin, log}$. In others words, the parametrisation is
\begin{equation}
    \delta P = A^{\rm X}_{\sin} \sin(\omega_{X} k) + A^{\rm X}_{\cos} \cos(\omega_{X} k)\;.
\end{equation}
The equations relating $(A_{\rm X}, \phi)$ with $(A^{\rm X}_{\sin}, A^{\rm X}_{\cos})$ are
\begin{eqnarray}
\label{eqn: sin_cos_to_phase_1}
A_{X}^2 &=& (A^{\rm X}_{\sin})^2 + (A^{\rm X}_{\cos})^2\;,\\
\label{eqn: sin_cos_to_phase_2}
\tan{\phi} &=& \frac{A^{\rm X}_{\cos}}{A^{\rm X}_{\sin}}\;.
\end{eqnarray}

\begin{figure}[t!]
    \centering
    \includegraphics[width = 0.5\textwidth]{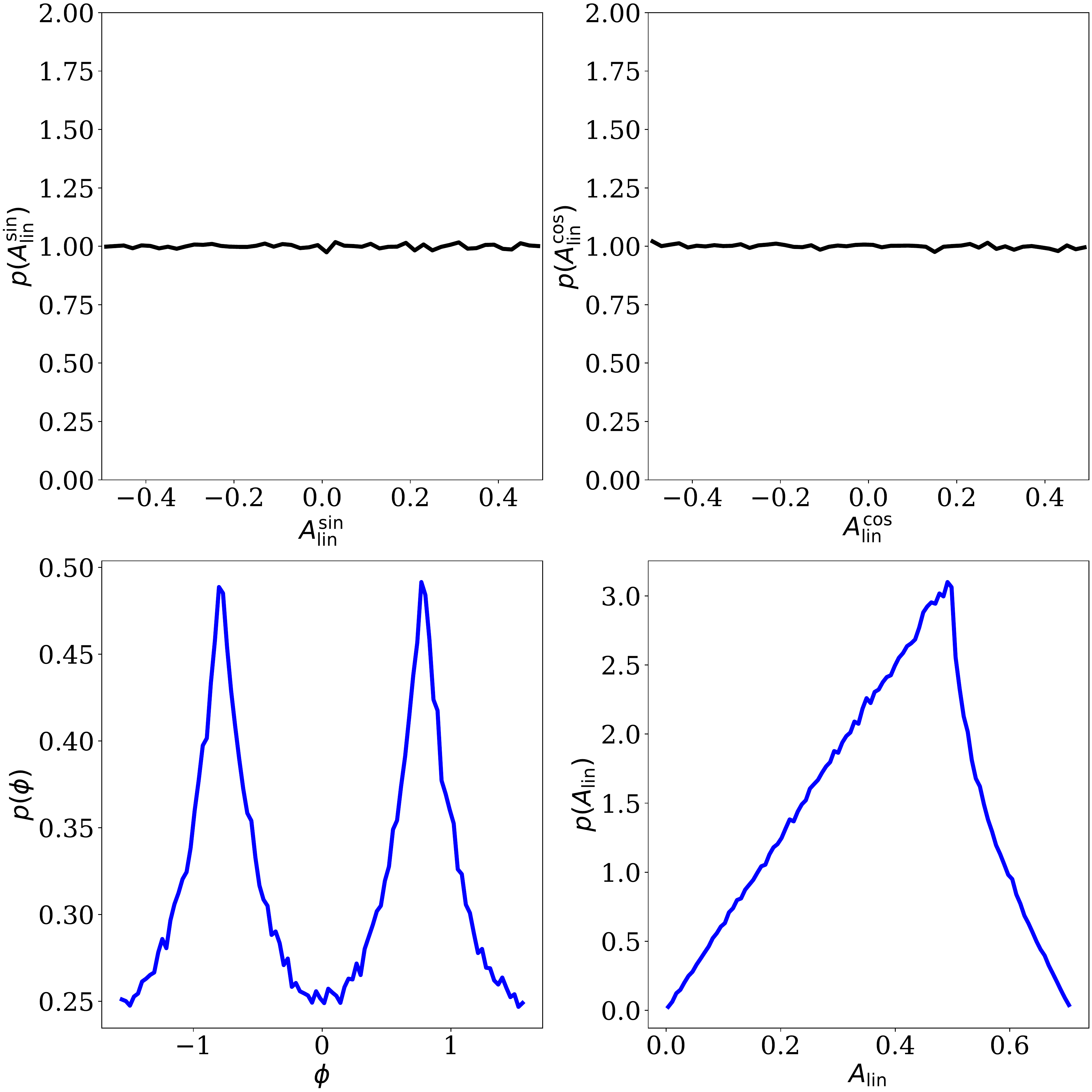}
    \includegraphics[width = 0.9\textwidth]{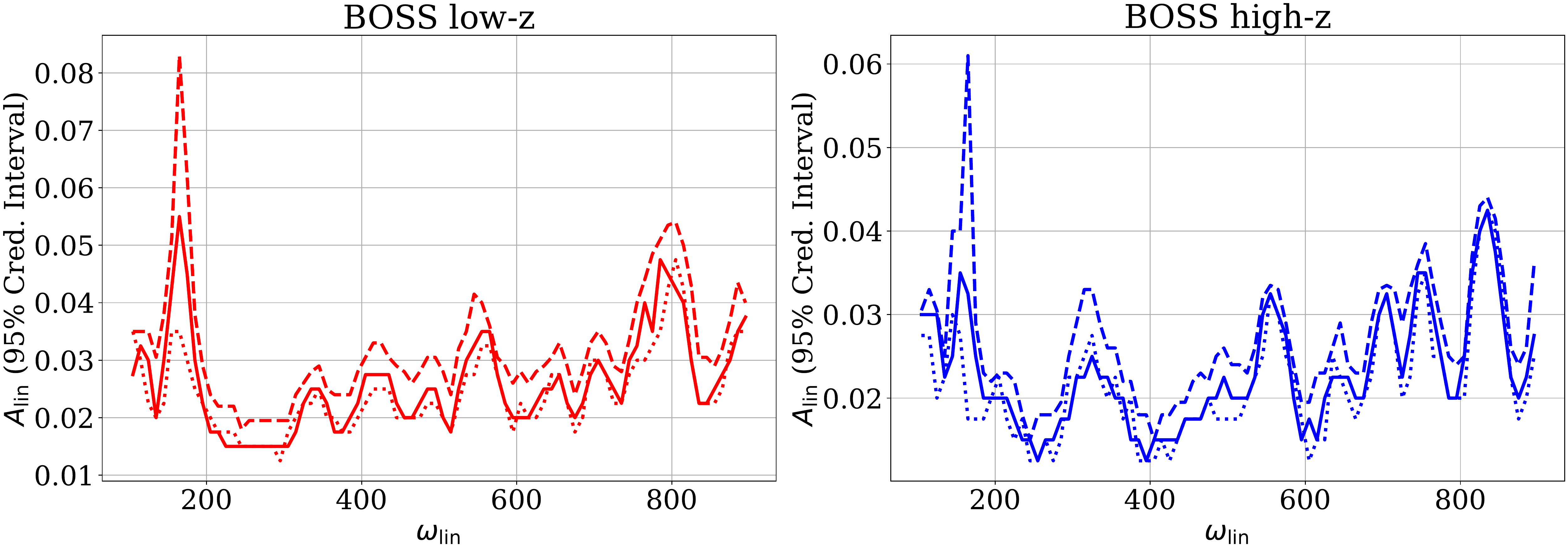}
    \caption{\textit{upper  panel}: Comparing the 95\% credible intervals on $A_{\rm lin}$ using $(A^{\rm lin}_{\rm sin}, A^{\rm lin}_{\rm cos})$ and the $(A_{\rm lin}, \phi)$ parametrisation. These two parametrisations are connected by a non-linear transformation, so to impose a uniform prior in one of them leads to a non-uniform prior in the other. \textit{bottom panel}: How the prior $\mathcal{U}(-0.5, 0.5)$ for $A^{\rm sin}_{\rm lin}$ and $A^{\rm cos}_{\rm lin}$ are changed when transformed into the $(A_{\rm lin}, \phi)$ parametrisation.}
    \label{fig:comparing_priors}
\end{figure}
\begin{figure}[t!]
    \centering
    \includegraphics[width = \textwidth]{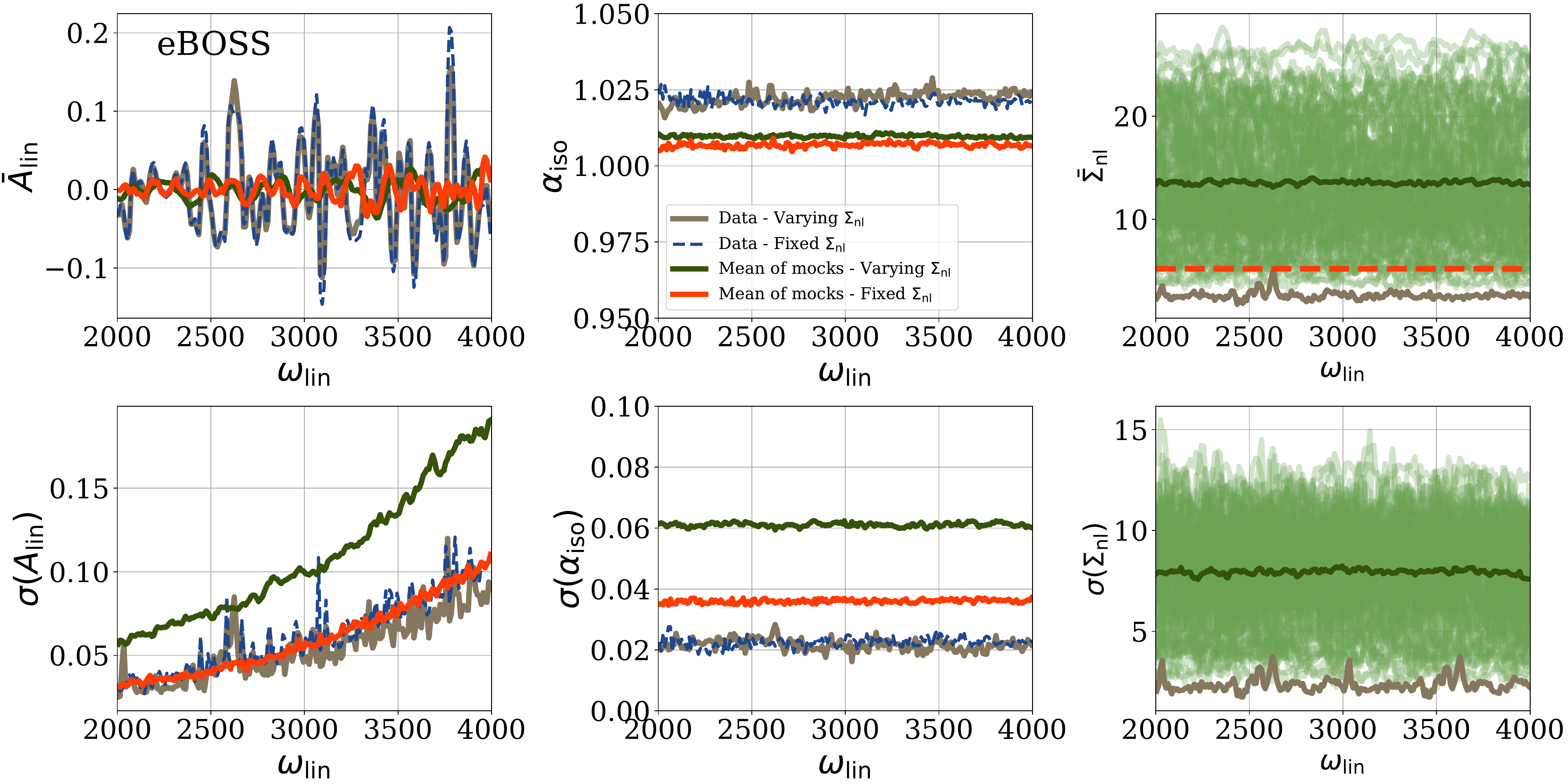}
    \includegraphics[width = \textwidth]{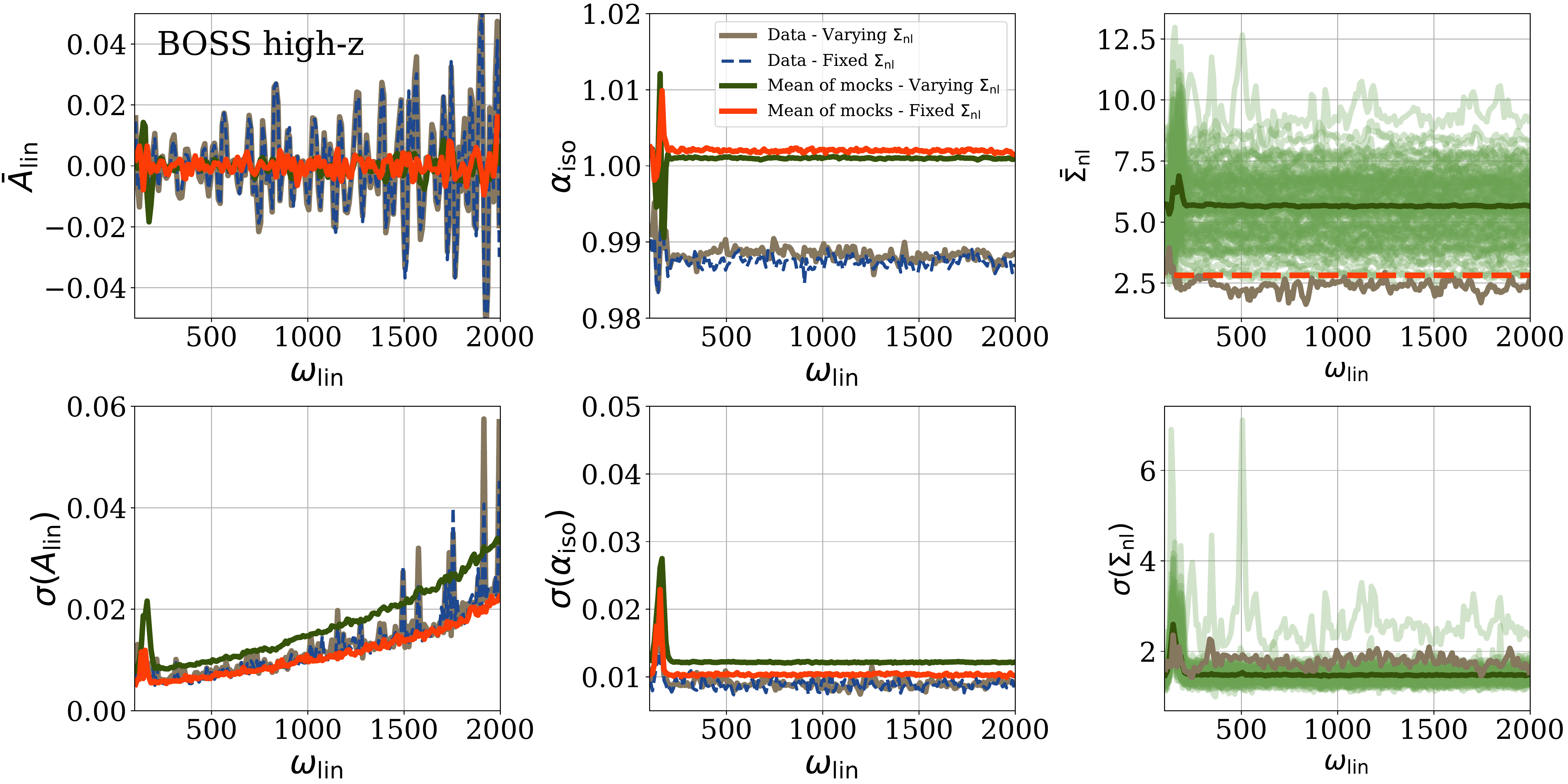}
    \caption{The influence of $\Sigma_{\rm nl}$ on linear feature constraints from the eBOSS Quasar sample and from the BOSS high-$z$ galaxy sample. The value of $\Sigma_{\rm nl}$ is fixed to the same value used on the official BAO analysis, \cite{Neveux:2020voa, BOSS:2016hvq}.}
    \label{fig: non_linear_impact}
\end{figure}
Since the equations above are non-linear, one should be careful when assigning the priors. A non-informative prior on the oscillations means that we have no knowledge about its amplitude and phase, so the natural choice for a non-informative prior would be, for instance, $p(A_{\rm X}) = \mathcal{U}(-0.5, 0.5)$ and $p(\phi) = \mathcal{U}(- \pi, \pi)$. According to the equations above, it is evident that these priors will be non-uniform in the parametrisation $(A^{\rm X}_{\sin}, A^{\rm X}_{\cos})$. In \cite{beutler2019primordial}, the authors decided to impose a uniform prior on $(A^{\rm X}_{\sin}, A^{\rm X}_{\cos})$, but according to the discussion above they are non-uniform on $(A_{\rm X}, \phi)$, and therefore informative. In the upper panel of Figure \ref{fig:comparing_priors} we show how a uniform prior on $A_{\rm sin}$ and $A_{\rm cos}$ is transformed when mapped into the phase parametrisation. To check the consistency of using an uniform prior on $(A^{\rm X}_{\sin}, A^{\rm X}_{\cos})$, we did the following. We first performed the analysis using the BOSS data and assuming $p(A^{\rm \rm lin}_{\sin}, A^{\rm lin}_{\cos}) = \mathcal{U}(-0.5, 0.5)$. The 95\% credible intervals for $A^{\rm lin}$ are represented by the solid lines in Figure \ref{fig:comparing_priors}. Next, we performed another analysis considering the phase parametrisation and assuming $p(A_{\rm lin}) = \mathcal{U}(-0.5, 0.5)$ and $p(\rm \phi) = \mathcal{U}(-\pi, \pi)$ and the results are represented by the dashed lines in the same figure. The credible intervals for the uniform parametrisation on $(A^{\rm X}_{\sin}, A^{\rm X}_{\cos})$ are more stringent, indicating that it's prior is slightly informative. Finally, we transformed the priors $p(A^{\rm \rm lin}_{\sin}, A^{\rm lin}_{\cos}) = \mathcal{U}(-0.5, 0.5)$ into priors on $(A_{\rm lin}, \phi)$ using Eqs.\,\ref{eqn: sin_cos_to_phase_1} and \ref{eqn: sin_cos_to_phase_2}, and the results are represented by the dotted lines in the bottom panel of Figure \ref{fig:comparing_priors}. We conclude that the most conservative choice is to use the parametrisation $(A_{\rm lin}, \phi)$.

\section{The impact of non-linear clustering}

\label{app: non-linear impact}
In Section \ref{sec:results} we showed that the mock constraints on global oscillatory features are noisier than the data. Here we show some further tests we performed to understand what may be causing this.

The main purpose of the mock catalogues produced for the BOSS and eBOSS datasets is to compute covariance matrices. Since performing high-resolution simulations is computationally costly and the number of catalogues needed to have reliable covariance matrix estimations is high, of the order of $\sim \mathcal{O}(10^3)$ mocks, it is necessary to simplify the process of catalogue generation. The mocks we used in this paper were generated by assuming approximate gravity solvers to obtain the large-scale dark matter field  \cite{Kitaura:2013cwa, Chuang:2014vfa}. To obtain the BOSS Patchy mocks the dark-matter particles were displaced from Lagrangian to Eulerian positions using second order Lagrangian perturbation theory (2LPT) \cite{Kitaura:2015uqa}. To obtain the EZmocks for eBOSS \cite{Zhao:2020bib}, on the other hand, the effective Zel’dovich approximation was assumed (i.e. linear Lagrangian perturbation theory). The tracers are introduced later on top of the matter field following a biasing scheme, along with the systematic effects inherent in each dataset \footnote{Some of these systematic effects include  survey geometry and veto masks, radial selection, fibre collision and redshift assignment failures (e.g. see \cite{BOSS:2016apd, Laurent:2017gze}).}. Both methods were widely tested and generated catalogues precise enough for estimating the covariance matrix for data analysis in cosmology \cite{Chuang:2014toa, Blot:2018oxk, Lippich:2018wrx}. Nevertheless, as we pointed out before, the whole pipeline for mock generation requires a series of approximations in both its theoretical and processing aspect. For instance, 2LPT does not completely capture the cosmic web dynamics up to the scales used to generate the models. Moreover, the resolutions of the density fields are low ($\sim 5 h^{-1}\;\mathrm{Mpc}$ for eBOSS EZmocks and $\sim 2.5 h^{-1}\;\mathrm{Mpc}$ for BOSS Patchy mocks), meaning that below these scales the tracers are positioned following a cloud-in-cell distribution. These and other limitations entail catalogues that although good enough for  estimating the covariance matrix, do not truthfully reproduce all ingredients present in the data (see \cite{Pellejero-Ibanez:2019lhz, Kitaura:2015uqa, Zhao:2020bib}). One consequence is that the BAO signal in these mocks is weaker than in data \cite{BOSS:2016hvq, Neveux:2020voa}. Another consequence, and this is the most relevant for constraining primordial features, is the fact these mocks also favour larger non-linear damping, $\Sigma_{\rm nl}$, due to the smooth galaxy distribution inside grid cells. This is emphasized in Figure \ref{fig: non_linear_impact}. We show the results for linear features, the isotropic BAO parameter, $\alpha_{\rm iso}$, and the non-linear damping parameter for both eBOSS and BOSS high-$z$ products as a function of the feature frequency. The solid green line represent the mean of the mocks with varying $\Sigma_{\rm nl}$, whereas the solid red line is the same but with the damping parameter fixed to the value used in the official BAO analysis of each dataset. For the amplitude of linear feature, fixing $\Sigma_{\rm nl}$ improves the mock constraint by a factor of $\sim 1.5$, leading to the same error as predicted by the data. The impact of fixing $\Sigma_{\rm nl}$ when analysing the data is not important because the value we obtain by varying it is the same as the value we fix. For both datasets, on the right, we show the scatter of $\Sigma_{\rm nl}$ for different mocks and their mean. Notice that in both cases none of the mocks have a $\Sigma_{\rm nl}$ compatible with the data.

We finish this section by emphasizing that having mock catalogues with features in the primordial power spectrum is crucial for learning how these signals can be measured in realistic scenarios and to understand how systematic effects can interfere with this process. Our findings show that the usual approximate methods to obtain numerically cheap mock catalogues may not be ideal to accomplish this, which requires more precise gravity solvers or full N-body simulations.

\bibliographystyle{JHEP}
\bibliography{main}

\end{document}